\documentclass{article}  

\usepackage{amsmath,amssymb,amsfonts}
\newtheorem{theorem}{Theorem}[section]
\newtheorem{lemma}[theorem]{Lemma}
\newtheorem{definition}[theorem]{Definition}
\newtheorem{corollary}[theorem]{Corollary}

\title{Asymptotically Flat Initial Data with Prescribed Regularity
at Infinity}
\author{Sergio Dain,\,\,Helmut Friedrich\\ 
Max-Planck-Institut f\"ur Gravitationsphysik\\
Am M\"uhlenberg 1\\
14476 Golm\\
Germany}

\begin{document}
\maketitle

\begin{abstract}
We prove the existence of a large class of asymptotically flat initial
data  with non-vanishing mass and angular momentum for which the metric
and the extrinsic curvature have asymptotic expansions at space-like 
infinity in terms of powers of a radial coordinate.
\end{abstract}

\newpage

\section{Introduction}

An initial data set for the Einstein vacuum equations is given by a
triple  $(\tilde S, \tilde h_{ab}, \tilde \Psi_{ab})$  where $\tilde S$
is a connected 3-dimensional manifold, $\tilde h_{ab} $ a (positive
definite) Riemannian metric, and $\tilde \Psi_{ab}$ a symmetric tensor
field on 
$\tilde S$. The data will be called  ``asymptotically flat'', if the
complement of a compact set in $\tilde S$ can be mapped by a coordinate
system $\tilde x^j$ diffeomorphically onto the complement of a closed
ball in $\mathbb{R}^3$ such that we have in these coordinates
\begin{equation} 
\label{pf1}
\tilde h_{ij}=(1+\frac{2m}{\tilde r})\delta_{ij}+O(\tilde r^{-2}),
\end{equation}
\begin{equation} 
\label{pf2}
\tilde \Psi_{ij}=O(\tilde r^{-2}),
\end{equation}
as $\tilde r= ( \sum_{j=1}^3 ({\tilde x^j})^2 ) ^{1/2} \to \infty$.
Here the constant $m$ denotes the mass of the data, $a,b,c...$ denote
abstract indices, $i,j,k...$, which take values $1, 2, 3$, denote
coordinates indices while $\delta_{ij}$ denotes the flat metric with respect
to the given coordinate system $\tilde x^j$. Tensor indices will be moved
with the metric $h_{ab}$ and its inverse $h^{ab}$. We set $x_i = x^i$ and
$\partial^i = \partial_i$. Our conditions guarantee that the mass, the
momentum, and the angular momentum of the initial data set are well defined. 

There exist  weaker notions of  asymptotic flatness (cf. \cite{Choquet81})
but they are not useful for our present purpose. In this article we show
the existence of a class of asymptotically flat initial data which have a
more controlled asymptotic behavior than (\ref{pf1}), (\ref{pf2}) in the
sense that they admit near space-like infinity asymptotic expansions of the
form 
\begin{equation} 
\label{he1}
\tilde h_{ij}\sim (1+\frac{2m}{\tilde r})\delta_{ij}+\sum_{k \geq2} 
\frac{\tilde h^k_{ij}}{\tilde r^k}
\end{equation}
\begin{equation} 
\label{Psie2}
\tilde \Psi_{ij}\sim \sum_{k \geq2} \frac{\tilde \Psi^k_{ij}}{\tilde r^k},
\end{equation}
where $\tilde h^k_{ij}$ and $\tilde \Psi^k_{ij}$ are smooth function on the 
unit 2-sphere (thought as being pulled back to the spheres 
$\tilde{r} = const.$ under the map 
$\tilde{x}^j \rightarrow \tilde{x}^j/\tilde{r}$).

We are interested in such data for two reasons. In \cite{Friedrich98} has
been studied in considerable detail the evolution of asymptotically flat
initial data near space-like and null infinity. There has been derived in
particular a certain ``regularity condition'' on the data near
space-like infinity, which is expected to provide a criterion for the
existence of a smooth asymptotic structure at null infinity. To simplify
the lengthy calculations, the data considered in \cite{Friedrich98} have
been assumed to be time-symmetric and to admit a smooth conformal
compactification. With these assumptions the regularity condition is given
by a surprisingly succinct expression.  With the present work we want to
provide data which will allow us to perform the analysis of
\cite{Friedrich98} without the assumption of time symmetry but which are
still ``simple'' enough to simplify the work of generalizing the regularity
condition to the case of non-trivial second fundamental form. 

Thus we will insist in the present paper on the smooth conformal
compactification of the metric but drop the time symmetry. A subsequent
article will be devoted to the analysis of a class of more general data
which will include in particular stationary asymptotically flat data.

The ``regular finite initial value problem near space-like infinity'',
formulated and analyzed in \cite{Friedrich98}, suggests to calculate
numerically entire asymptotically flat solutions to Einstein's vacuum field
equations on finite grids. In the present article we provide data for such
numerical calculations which should allows us to study interesting
situations while keeping a certain simplicity in the handling of the
initial data.   

The difficulty of constructing data with the asymptotic behavior 
(\ref{he1}), (\ref{Psie2}) arises from the fact that the fields   
need to satisfy the constraint equations 
\[
\tilde D^b \tilde \Psi_{ab} -\tilde D_a \tilde \Psi=0
\]
\[
\tilde R + \tilde \Psi^2-\tilde \Psi_{ab} \tilde \Psi^{ab}=0,
\]
on $\tilde S$, where $\tilde D_a$ is the covariant derivative, $\tilde R$ is
the trace of the corresponding Ricci tensor, and 
$\tilde \Psi=\tilde h^{ab} \tilde \Psi_{ab}$. Part of the data, the ``free
data'', can be given such that (\ref{he1}), (\ref{Psie2}) hold. However, the
remaining data are governed by elliptic equations and we have to show that
(\ref{he1}), (\ref{Psie2}) are in fact a consequence of the equations and
the way the free data have been prescribed.   

To employ the known techniques to provide solutions to the constraints, 
we assume   
\begin{equation}
\label{physmax}
\tilde \Psi = 0, 
\end{equation}
such that the data correspond to a hypersurface which is maximal in the
solution space-time.

We give an outline of our results. Because of the applications
indicated above, we wish to control in detail the conformal structure of
the data near space-like infinity. Therefore we shall analyze the data
in terms of the conformal compactification $(S, h_{ab}, \Psi_{ab})$
of the ``physical'' asymptotically flat data. Here $S$ denotes a smooth,
connected, orientable, compact 3-manifold. It contains a point $i$ such
that we can write $\tilde{S} = S\backslash \{ i \} $.  The point $i$ will
represent, in a sense described in detail below, space like infinity for
the physical initial data. 

By singling out more points in $S$ and by treating the
fields near these points in the same way as near $i$ we could construct
data with several asymptotically flat ends, since all the following
arguments equally apply to such situations. However, for convenience we
restrict ourselves to the case of a single asymptotically flat end.   

We assume that $h_{ab}$ is a positive definite metric on $S$ with covariant
derivative $D_a$ and $\Psi_{ab}$ is a symmetric tensor field which is
smooth on $\tilde{S}$. In agreement with (\ref{physmax}) we shall assume
that
$\Psi_{ab}$ is trace free, 
\[
h^{ab}\,\Psi_{ab} = 0. 
\]
The fields above are related to the physical fields by rescaling
\begin{equation}
\label{hPrescale}
\tilde{h}_{ab} = \theta^4\,h_{ab},\,\,\,\,\,\,\, 
\tilde{\Psi}_{ab} = \theta^{-2}\,\Psi_{ab},
\end{equation}
with a conformal factor $\theta$ which is positive on $\tilde{S}$.
For the physical fields to satisfy the vacuum constraints we need to assume
that
\begin{equation} 
\label{diver}
D^a \Psi_{ab}=0 \quad\mbox{on}\quad \tilde{S},
\end{equation}
\begin{equation} 
\label{Lich}
(D_{b}D^{b}-\frac{1}{8}R)\theta=-\frac{1}{8}\Psi_{ab}\Psi^{ab}\theta^{-7}
\quad\mbox{on}\quad \tilde{S}. 
\end{equation}
Equation (\ref{Lich}) for the conformal factor $\theta$ is the Lichnerowicz
equation, transferred to our context. 

Let $x^j$ be $h$-normal coordinates centered at $i$ such that 
$h_{kl}= \delta_{kl}$ at $i$ and set $r = (\sum_{i=1}^3 (x^j)^2 ) ^{1/2}$. 
To ensure asymptotical flatness of the data (\ref{hPrescale}) we require   
\begin{equation} 
\label{Psii}
\Psi_{ab} = O(r^{-4}) \quad\mbox{as}\quad r \rightarrow 0,  
\end{equation}
\begin{equation} 
\label{thetai}
\lim_{r\to 0}r\theta = 1.
\end{equation}
In the coordinates $\tilde{x}^j = x^j/r^2$ the fields (\ref{hPrescale})
will then satisfy (\ref{pf1}), (\ref{pf2}) (cf. \cite{Friedrich88},
\cite{Friedrich98} for this procedure).

Not all data (\ref{hPrescale}) derived from data $h_{ab}$, $\Psi_{ab}$
as described above will satisfy conditions (\ref{he1}), (\ref{Psie2}). 
We will have to impose extra conditions and we want to keep these conditions
as simple as possible.  

Since we assume the metric $h_{ab}$ to be smooth on $S$, it will
only depend on the behavior of $\theta$ near $i$ whether condition
(\ref{he1}) will be satisfied. Via equation (\ref{Lich}) this behavior
depends on $\Psi_{ab}$.  What kind of condition do we have to impose on
$\Psi_{ab}$ in order to achieve (\ref{he1}) ?  

The following space of functions will play an important role in our
discussion. Denote by $B_a$ the open ball with
center $i$ and radius $r = a > 0$, where $a$ is chosen small enough such
that $B_a$ is a convex normal neighborhood of $i$.   
A function $f\in C^\infty(\tilde S)$ is said to be in $E^\infty(B_a)$
if on  $B_a$ we can write $f = f_1 +rf_2 $
with $f_1, f_2 \in C^\infty(B_a)$ (cf. definition \ref{dEm}).

An  answer to our question is given by following theorem. 
\begin{theorem} 
\label{T1}
Let $h_{ab}$ be a smooth metric on $S$ with positive Ricci scalar $R$. 
Assume  that $\Psi_{ab}$ is smooth in $\tilde S$ and satisfies on
$B_a$   
\begin{equation} 
\label{condd}
r^8\Psi_{ab} \Psi^{ab}\in E^\infty(B_a).
\end{equation} 
Then there exists  on $\tilde S$ a unique solution $\theta$ of 
equation (\ref{Lich}), which is positive, satisfies (\ref{thetai}), 
and has in $B_a$ the form
\begin{equation} 
\label{1f}
\theta = \frac{\hat \theta}{r}, \quad \hat \theta \in E^\infty(B_a),
\quad \hat \theta (i)=1.
\end{equation}
\end{theorem}

In fact, we will get slightly more detailed information. We find that 
$\hat \theta= u_1 + r\,u_2$ on $B_a$ with $u_2 \in E^\infty(B_a)$ and a
function $u_1\in C^\infty(B_a)$ which satisfies $u_1=1+O(r^2)$ and
\[
(D_{b}D^{b}-\frac{1}{8}R)\frac{u_1}{r}= \theta_R, 
\]
in  $B_a \backslash \{i\}$, where $\theta_R$ is in $ C^\infty(B_a)$
and vanishes at any order at $i$. 

If $\theta$ has the form (\ref{1f}) then  (\ref{he1}) will be satisfied
due to our assumptions on $h_{ab}$.  

Note the simplicity of condition (\ref{condd}). To allow for later
generalizations, we shall discuss below the existence of the solution
$\theta$ under weaker assumptions on the smoothness of the metric
$h_{ab}$ and the smoothness and asymptotic behavior of $\Psi_{ab}$ 
(cf. theorem \ref{Beig}). In fact, already the methods used in
this article would allow us to deduce analogues of all our results under
weaker differentiability assumptions, however, we are particularly
interested in the $C^{\infty}$ case because it will be convenient in our
intended applications.  If the metric is analytic on $B_a$ it can be
arranged that $\theta_R = 0$ and $u_1$ is analytic on $B_a$ (and unique
with this property, see \cite{Garabedian} and the remark after Theorem
\ref{Green}). We finally note that the requirement $R > 0$, which ensures
the solvability of the Lichnerowicz equation, could be reformulated in
terms of a condition on the Yamabe number (cf. \cite{Lee87}).

It remains to be shown that condition (\ref{condd}) can be satisfied 
by tensor fields $\Psi_{ab}$ which satisfy (\ref{diver}), (\ref{Psii}). A
special class of such solutions, namely those which extend smoothly to all
of $S$, can easily be obtained by known techniques (cf. \cite{Choquet80}).
However, in that case the initial data will have vanishing momentum and
angular momentum. To obtain data without this restriction, we have to
consider fields $\Psi_{ab} \in C^{\infty}(\tilde{S})$ which are singular at
$i$ in the sense that they admit, in accordance with (\ref{pf2}),
(\ref{hPrescale}), (\ref{thetai}), at $i = \{r = 0\}$ asymptotic expansions 
of the form
\begin{equation} 
\label{PsiS}
\Psi_{ij}\sim \sum_{k\geq -4} \Psi^k_{ij} r^k
\quad\mbox{with}\quad \Psi^k_{ij} \in C^{\infty}(S^2). 
\end{equation}

It turns out that condition (\ref{condd}) excludes data with 
non-vanishing linear momentum, which requires a non-vanishing leading order
term of the form $O(r^{-4})$. In section \ref{logs} we will show that such
terms imply terms of the form $\log\,r$ in $\theta$  and thus do not
admit expansion of the form (\ref{he1}). However, this does not
necessarily indicate that condition (\ref{condd}) is overly restrictive. In
the case where the metric $h_{ab}$ is smooth  it will be shown in section
\ref{logs}  that a non-vanishing linear momentum always comes with
logarithmic terms, irrespective of whether condition (\ref{condd}) is
imposed or not.    

There remains the question whether there exist fields $\Psi_{ab}$ which
satisfy  (\ref{condd}) and have non-trivial angular momentum. The
latter requires a term of the form $O(r^{-3})$ in (\ref{PsiS}). It turns
out that condition (\ref{condd}) fixes this term to be of the form
\begin{equation} 
\label{PsiJ}
\Psi^{AJ}_{ij}=\frac{A}{r^3}  (3n_in_j - \delta_{ij})
+\frac{3}{r^3}(n_j  \epsilon_{kil} J^l n^k +  n_i\epsilon_{ljk} J^k n^l), 
\end{equation}
where $n^i = x^i/r$ is the radial unit normal vector field near $i$ and
$J^k$, $A$  are constants, the three constants $J^k$ specifying the
angular momentum of the data. The spherically symmetric tensor which
appears here with the factor $A$ agrees with the extrinsic curvature for a
maximal (non-time symmetric) slice in the Schwarzschild solution (see for
example \cite{Beig98}). Note that the tensor $\Psi^{AJ}_{ij}$ satisfies
condition (\ref{condd}) and the equation
$\partial^i \Psi_{ij}^{AJ}=0$ on $\tilde S$ for the flat metric. In the 
next theorem we prove an analogous result for general smooth metrics. 

\begin{theorem} 
\label{T2}
Let $h_{ab}$  a smooth metric in $S$. There exists trace-free tensor 
fields $\Psi_{ab} \in C^{\infty}(S \setminus \{i\})$ satisfying
(\ref{PsiS}) with the following properties

(i) $\Psi_{ab}=\Psi^{AJ}_{ab}+\hat\Psi_{ab}$, where $\Psi^{AJ}_{ab}$ is
given  by (\ref{PsiJ}) and $\hat\Psi_{ab}=O(r^{-2})$.

(ii) $D^a\Psi_{ab}=0$ on $\tilde S$

(iii) $r^8 \Psi_{ab}\Psi^{ab}$ satisfies condition (\ref{condd}).
\end{theorem}

We prove a more detailed version of this theorem in section 
(\ref{asymtoticmomentum}). There it will be shown how to construct such 
solutions from free-data by using the York splitting technique
(\cite{York73}). In section \ref{flatmomentum} the case where $h_{ab}$ is
conformal to the Euclidean metric is studied in all generality.

\section{Preliminaries}
\label{preliminaries}

In this section we collect some known facts from functional analysis and
the theory of linear elliptic partial differential equations.

Let $\mathbb{Z}$ be the set of integer numbers and  $\mathbb{N}_0$  the 
set of non negative integers. We use multi-indices 
$\beta=(\beta_1, \beta_2, \dots, \beta_n) \in \mathbb{N}_0^n$ 
and set $|\beta|=\sum_{i=1}^n \beta_i$,
$\beta ! = \beta_1 ! \beta_2 ! \dots \beta_n !$, 
$x^{\beta} = (x^1)^{\beta_1} (x^2)^{\beta_2} \dots (x^n)^{\beta_n}$,
$\partial^\beta u = \partial_{x^1}^{\beta_1}\,\partial_{x^2}^{\beta_2}
\ldots \partial_{x^n}^{\beta_n}\,u$, 
$D^\beta u = D_{1}^{\beta_1}\,D_{2}^{\beta_2}\ldots
D_n^{\beta_n}\,u$, and, for $\beta, \gamma \in \mathbb{N}_0^n$, 
$\beta + \gamma = (\beta_1 + \gamma_1, \ldots, \beta_n + \gamma_n)$
and $\beta \le \gamma$ if $\beta_i \le \gamma_i$. We denote by $\Omega$
an open domain in $\mathbb{R}^3$ (resp. in $S$; quite often we will then
choose
$\Omega = B_a$).

We shall use the following functions spaces (see \cite{Adams},
\cite{Gilbarg} for  definitions, notations, and results): the set of $m$
times continously differentiable functions $C^m(\Omega)$, the H\"older space
$C^{m, \alpha}(\Omega)$, where $0<\alpha<1$, the corresponding
spaces $C^m(\bar \Omega)$, $C^{m, \alpha}(\bar \Omega)$, the space
$C^\infty_0(\Omega)$  of smooth function with compact support in
$\Omega$, the Lebesgue space $L^p(\Omega)$, the Sobolev space
$W^{m,p}(\Omega)$, and the local Sobolev space $W^{m,p}_{loc}(\Omega)$.
For a compact manifold $S$ we can also define analogous spaces $L^p(S)$, 
$C^{m, \alpha}(S)$, $W^{m,p}(S)$ (cf. \cite{Aubin82}). 

We shall need the following relations between these spaces. 

\begin{theorem}[Sobolev imbedding] 
\label{imbedding}
Let $\Omega$ be a $C^{0,1}$ domain in $\mathbb{R}^3$, let  $k$, $m$,
$j$ be  
non-negative integers and $1\leq p,q<\infty$. Then there exist 
the following imbeddings:

(i) If $mp<3$, then 
\[
W^{j+m,p}(\Omega) \subset W^{j,q}(\Omega), \quad p\leq q\leq 3p/(3-mp).
\]

(ii) If $(m-1)p < 3 < mp$, then   
\[
W^{j+m,p}(\Omega) \subset C^{j,\alpha}(\bar \Omega), \quad \alpha=m-3/p.
\]
\end{theorem}

\begin{theorem} 
\label{Morrey}
Let $u\in W^{1,1}(\Omega)$, and suppose there exist positive constants 
$\alpha \leq 1$ and  $K$ such that
\[
\int_{B_R} |\partial u| \, d \mu \leq K R^{2+\alpha} \quad \text{for all 
balls} \quad B_R \subset \Omega 
\quad \text{of radius} \quad R > 0.
\]
Then $u\in C^\alpha(\Omega)$.
\end{theorem}

Our existences proof for the non-linear equations relies on the following 
version of the compact imbedding for compact manifolds \cite{Aubin82}.  
\begin{theorem}[Rellich-Kondrakov] 
\label{Rellich}
The following imbeddings are compact

(i) $W^{m,q}(S) \subset L^p(S)$ if  $1 \geq 1/p >1/q-m/3>0$

(ii) $W^{m,q}(S) \subset C^\alpha(S)$ if $m-\alpha>3/q$,  
$0\leq\alpha<1$. 
\end{theorem}
A further essential tool for the existence proof is the Schauder fixed
point theorem \cite{Gilbarg}.
\begin{theorem}[Schauder fixed point] 
\label{Schauder}
 Let $B$ be a closed convex set in a Banach space $V$ and let $T$ be a 
continuous mapping of $B$ into itself such that the image $T(B)$ is
precompact, i.e. has compact closure in $B$. Then $T$ has a fixed point.
\end{theorem}

We turn now to the theory of elliptic partial differential equations 
(see \cite{Cantor}, \cite{Choquet81}, \cite{Gilbarg}, \cite{Morrey66}).
Let $\mathbf{L}$ be a linear differential operator of order $m$ on the
compact manifold $S$ which acts on tensors fields $u$. In the case where 
$u \sim u^{a_1 \ldots a_{m_1}}$ is a contravariant tensor field
of rank $m_1$, $\textbf{L}$ has in local coordinates the form
\begin{equation} \label{eq:elliptic}
\mathbf{L}u = 
\sum_{|\beta|=0}^m a^{j_1 \ldots j_{m_2}}\,    
_{i_1 \ldots i_{m_1} \beta}\,D^\beta\,
u^{i_1 \ldots i_{m_1}}\equiv 
\sum_{|\beta|=0}^m a_\beta \, D^\beta\,
u,
\end{equation}
where the  coefficients $a^{j_1 \ldots j_{m_2}}\,_{i_1 \ldots i_{m_1} \beta} =
a(x)^{j_1 \ldots j_{m_2}}\,_{i_1 \ldots i_{m_1} \beta}$ are tensor
fields  of a certain
smoothness, and $D$ denotes the Levi-Civita connexion with respect to
the metric $h$. In the expression on the right-hand side we suppressed the
indices belonging to the unknown and the target space. 
Assuming the same coordinates as above, we write for given
covector $\xi_i$ at a point $x \in \Omega$ and multi-index $\beta$ as usual 
$\xi^{\beta} = \xi_1^{\beta_1} \ldots \xi_4^{\beta_4}$ and define 
a linear map $A(x, \xi): \mathbb{R}^{m_1} \rightarrow \mathbb{R}^{m_2}$
by setting $(A(x, \xi)\,u)^{j_1 \ldots j_{m_2}} =
\sum_{|\beta|=m} a(x)^{j_1 \ldots j_{m_2}}\,_{i_1 \ldots i_{m_1}\,\beta} 
\,\xi^{\beta}\,u^{i_1 \ldots i_{m_1}}$.
The operator $\mathbf{L}$ is elliptic at $x$ if for any $\xi \neq 0$ the
map $A(x, \xi)$ is an isomorphism, $\mathbf{L}$ is elliptic on $S$ if it
is elliptic at all points of $S$. 

We have the following $L^p$ regularity result
\cite{Agmon59}, \cite{Agmon64}, \cite{Choquet81}, \cite{Morrey66}. 

\begin{theorem}[$L^p$ regularity] 
\label{globalregularity}
Let $\mathbf{L}$ be an elliptic operator of order $m$ on $\Omega$ (resp.
$S$) with  
coefficients $a_\beta \in W^{s_{|\beta|}, p}(S)$, where $s_k >
3/p+k-m+1$, and $p>1$. Let $s$ be a  natural number such that $s_k\geq
s-m\geq 0$.  
Let $u \in W^{m,p}_{loc}(\Omega)$ (resp. $W^{m,p}_{loc}(S)$), with
$p > 1$, be a solution of the  elliptic equation $\mathbf{L}u=f$.

(i) If $f\in W^{s-m,q}_{loc}(\Omega)$, $q\geq p$, then $u \in 
W^{s,q}_{loc}(\Omega)$.

(ii) If $f\in W^{s-m,q}(S)$, $q\geq p$, then $u \in W^{s,q}(S)$.
\end{theorem}

Furthermore, we have the  Schauder interior elliptic regularity
\cite{Agmon64},
\cite{Douglis55}, \cite{Gilbarg}, \cite{Morrey66}.
\begin{theorem}[Schauder elliptic regularity] 
\label{SchauderEstimate}
Let  $\mathbf{L}$ be an elliptic operator of order $m$ on $\Omega$ with  
coefficients $a_\beta \in C^{k, \alpha}(\bar\Omega)$. Let $u \in W^{m,p}(\Omega)$, with
$p > 1$, be a solution of the  elliptic equation $\mathbf{L}u=f$, with 
$f\in C^{k, \alpha}(\bar\Omega)$. Then $u\in C^{k+m,\alpha}(\Omega')$, 
for all $\Omega' \subset \subset \Omega$.
\end{theorem}

For linear elliptic equations we have the Fredholm alternative for
elliptic operators on compact manifolds \cite{Cantor}.

\begin{theorem}[Fredholm alternative]
\label{Cantor}

Let $\mathbf{L}$ be an elliptic operator of order $m$ on $S$ whose   
coefficients satisfy the hypothesis of Theorem
\ref{globalregularity}. Let $s$ be some natural number such $s_k\geq
s-m\geq 0$ and $f\in L^p(S)$, $p > 1$. Then the equation $\mathbf{L}u=f$
has a solution $u\in W^{m,p}(S)$ iff
\[
\int_S <v,f>_h d \mu =0 \quad \text{for all} \quad v\in \ker (\mathbf{L}^*).
\]
\end{theorem}
Here $d\mu$ denotes the volume element determined by $h$ and
$\mathbf{L}^*$ the formal adjoint of $\mathbf{L}$, which for the operator
(\ref{eq:elliptic}) is given by
\begin{equation}\label{eq:adjoint}
\mathbf{L}^*u = \sum_{|\beta|=0}^m (-1)^{|\beta|} D^\beta(a_\beta 
u).  
\end{equation}
Furthermore  $<,>_h$ denotes  the appropriate inner product induced by the metric 
$h_{ab}$. In our case, where $u$ and $f$ will be vector fields $f^a$ and
$u^a$, we have $<u,f>_h \,=f^au_a$.

Let
\begin{equation}
 \label{LL}
Lu=\partial_i(a^{ij}\partial_ju+b^i u)+c^i \partial_i u+ du,
\end{equation}
be a linear elliptic operator of second order   with principal part in 
divergence form  on $\Omega$ which acts on scalar functions.  
An operator  of the form (\ref{LL}) may be written 
in the form (\ref{eq:elliptic}) provided its principal coefficients
$a^{ij}$ are differentiable. 

We shall assume that $L$  is
strictly elliptic in $\Omega$; that is,  there exists  $\lambda > 0$ such that
\begin{equation} \label{eq:strictly}
a^{ij}(x) \xi_i \xi_j \geq \lambda |\xi |^2, \quad \forall x \in \Omega, 
\quad \xi \in \mathbb{R}^n.
\end{equation}

We also assume that $L$ has bounded coefficients; that is for some
constants $\Lambda$ and $\nu \geq 0$ we have for all $x\in \Omega$
\begin{equation}
  \label{eq:boundedc}
  \sum |a^{ij}|^2 \leq \Lambda^2, \quad \lambda^{-2} \sum (|b^i|^2
  +|c^i|^2)+\lambda^{-1} |d| \leq \nu^2.
\end{equation}

In order to formulate the maximum principle, we have to impose that
the coefficient of $u$ satisfy the non-positivity condition
\begin{equation}
  \label{eq:negative}
  \int_\Omega (dv -b^i \partial_i v)\, dx\leq 0 \quad \forall v\geq
  0,\, v\in C^1_0(\Omega). 
\end{equation}

We have the following versions of the maximum principle \cite{Gilbarg}.
\begin{theorem}[Weak Maximum Principle] 
\label{weakmaximum}
Assume that $L$ given by (\ref{LL}) satisfies conditions
(\ref{eq:strictly}), (\ref{eq:boundedc}) and (\ref{eq:negative}).  
Let $u\in W^{1,2}(\Omega)$ satisfy $Lu \geq 0$ ($\leq 0$) in $\Omega$. Then 
\[
\sup_{\Omega} u \leq \sup_{\partial \Omega} u^+ \,\,\,\, 
(\inf_{\Omega} u \geq \inf_{\partial \Omega} u^-),
\]
where  $u^+(x) = \max\{u(x), 0\}$, $u^-(x) = \min\{u(x), 0\}$.
\end{theorem}
  
\begin{theorem}[Strong Maximum Principle] \label{strongmaximum}
Assume that $L$ given by (\ref{LL}) satisfies conditions
(\ref{eq:strictly}), (\ref{eq:boundedc}) and (\ref{eq:negative}). Let $u\in W^{1,2}(\Omega)$ satisfy $Lu \geq 0$ in $\Omega$. 
Then, if for some ball $B \subset \subset \Omega$ we have
\[
\sup_B u=\sup_\Omega u \geq 0,
\]
the function $u$ must be constant in $\Omega$.
\end{theorem}
Because $u$ is assumed to be only in $W^{1,2}$ the inequality 
$Lu \geq 0$ has to be understood in the weak sense (see \cite{Gilbarg} for
details). 

\section{The Hamiltonian Constraint}

In this section  we will proof theorem (\ref{T1}). 
\subsection{Existence} \label{hexistence}

The existence of solutions to the Lichnerowicz equation has been studied
under various assumptions (cf. \cite{Choquet99}, \cite{Choquet80},
\cite{Isenberg} and the reference given there). The setting outlined
above, where we have to solve (\ref{Lich}), (\ref{thetai}) on the compact
manifold $S$, has been studied in \cite{Beig}, \cite{Friedrich88},
\cite{Friedrich98}.  

In general the ``physical'' metric provided by an asymptotically
flat initial data set will not admit a smooth conformal
compactification at space-like infinity. Explicit examples for such
situations can be obtained by studying space-like slices of stationary
solutions like the Kerr solution. To allow for later generalizations of
the present work which would admit also stationary solutions we shall
prove the existence result of theorem \ref{T1} for metrics $h_{ab}$ which
are not necessarily smooth. The proof we will employ Sobolev spaces
$W^{m,p}(S)$ and the corresponding imbeddings and elliptic estimates 
(in particular, there will be no need for us to employ weighted Sobolev
spaces with weights involving the distance to the point $i$). 
With these spaces and standard $L^p$ elliptic theory we will also be
able to handle the mild $r^{-1}$-type singularity at $i$ which occurs on
the right hand side of equation (\ref{Lich}).

The  conformal Laplacian or Yamabe operator  
\[
L_h=h^{ab}D_{a}D_{b}-\frac{1}{8}R,
\]
which appears of the left hand side of (\ref{Lich}), is a linear
elliptic operator of second order whose coefficients depend
on the derivatives of the metric $h$ up to second order. 
The smoothness to be required of the metric $h$ is determined by the
following considerations. In the existence proof we need:

\begin{itemize}
\item[(i)] The existence of normal coordinates. This suggests to assume
  $h\in C^{1,1}(S)$.

\item[(ii)] The maximum principle, theorems \ref{weakmaximum} and
  \ref{strongmaximum}. The required boundedness of the Ricci scalar $R$
  imposes restrictions on the second derivative of $h$.

\item[(iii)] The elliptic $L^p$ estimate, theorem
  \ref{globalregularity}. This requires that 
$h\in W^{3,p}(S)$ for $p>3/2$.

\end{itemize}

Since the right-hand side of equation (\ref{Lich}) is in $L^2$ the
assumption that
$h\in W^{3,p}(S)$, $p>3$, would be sufficient to handle equation
(\ref{Lich}). However, when we will discuss the momentum
constraint in section \ref{existencemom}, we will wish to be able
to handle cases where $p<3/2$. In these cases the conditions of theorem
\ref{globalregularity} suggest to assume that $h\in  W^{4,p}(S)$. 
In order to simplify our hypothesis we shall assume in the following that 
\begin{equation}
\label{eq:sobh}
h_{ab} \in W^{4,p}(S), \quad p>3/2.  
\end{equation}
The imbedding theorems then imply that
$h_{ab} \in C^{2,\alpha}(S)$, $0<\alpha<1$, whence $R\in C^{\alpha}(S)$.
We note that (\ref{eq:sobh}) is not the weakest possible assumption but
it will be sufficient for our future applications. 

\begin{lemma} 
\label{existenceL}
Assume that $h$ satisfies (\ref{eq:sobh})  and $R >0$.  Then: 

\noindent
(i) $L_h:W^{2,q}(S)\rightarrow L^q(S)$, $q>1$,  defines an
isomorphism. 

\noindent
(ii) if $u \in W^{1,2}(S)$ and $L_h u \le 0$, then $u \ge 0$; if,
moreover, $L_h u \neq 0 \in L^q(S)$, then $u > 0$.
\end{lemma}

\noindent
{\bf Proof:} 
(i) To show injectivity, assume that $L_h u = 0$. By elliptic regularity
$u$ is smooth enough such that we can multiply this equation with $u$
and integrate by parts to obtain
\[
\int_S  \left( D^au D_a u + \frac{1}{8} R u^2 \right) \, d\mu_h=0.
\]
Since $R>0$ it follows that $u=0$. Surjectivity follows then by theorem
\ref{Cantor} since $L_h = L_h^*$. Boundedness of $L_h$ is immediately
implied by the assumptions while the inequality  
\[
||u||_{W^{2,p}(S)}\leq C ||L_hu||_{L^p(S)},
\]
which follows from the elliptic estimates underlying theorem
\ref{globalregularity} and the injectivity of $L_h$ 
(see e.g \cite{Cantor} for this well known result), implies the
boundedness of $L^{-1}_h$.  

(ii) If we have $u\leq 0$ in some region of $S$, it follows that 
$\sup_{S} (-u) \geq 0$. Then there is a region in $S$ in which we can
apply the maximum principle to the function $-u$
to conclude that $u$ must be a non-positive constant whence 
$L_hu = -Ru/8 \geq 0$ in that region. In the case where 
$L_h u < 0$ we would arrive at a contradiction. In the case where
$L_h u \le 0$ we conclude that $u = 0$ in the given region and a
repetition of the argument gives the desired result. 
$\blacksquare$\\

To construct an approximate solution we choose normal coordinates $x^j$ 
centered at $i$ such that (after a suitable choice of $a > 0$) we have
in the open ball $B_a$ in these coordinates
\begin{equation}
\label{hsplit}
h_{ij}=\delta_{ij}+\hat h_{ij},\,\,\,\,\,\,
h^{ij}=\delta^{ij}+\hat h^{ij}
\end{equation}
with
\[
\hat h_{ij}=O(r^2),\,\,\,\,\,\,\hat h^{ij}=O(r^2),\,\,\,\,\,\,
x^i \hat h_{ij}=0,\,\,\,\,\,\,x_i \hat h^{ij}=0.
\]
Notice that $\hat h_{ij}$, $\hat h^{ij}$, defined by the equations above,
are not necessarily related to each other by the usual process of raising
indices.

Denoting by $\Delta$ the flat Laplacian with respect to the coordinates
$x^j$, we write on $B_a$ 
\[
L_h= \Delta +\hat L_h,
\]
with
\begin{equation} 
\label{Lhath}
\hat L_h= \hat h^{ij}\partial_i \partial_j+b^i\partial_i-\frac{1}{8} R.
\end{equation}
We note that 
\[
b^i=O(r).
\]

Choose a function $\chi_a \in C^\infty (S)$ which is non-negative and
such that $\chi_a = 1$ in $B_{a/2}$ and $\chi_a = 0$ in 
$S \setminus B_a$. Denote by $\delta_i$ the Dirac delta distribution
with source at $i$. 

\begin{lemma} 
\label{existenceGreen}
Assume that $h$ satisfies (\ref{eq:sobh}) and $R>0$. Then, there
exists a unique solution $\theta_0$ of the equation 
$L_h \theta_0=-4\pi \delta_i$. Moreover $\theta_0>0$ in $\tilde S$ and we
can write $\theta_0 = \chi_a/r + g$ with $g\in C^\alpha(S)$,
$0<\alpha<1$. \end{lemma}

\noindent
{\bf Proof:} Observing that $1/r$ defines a fundamental solution to the
flat Laplacian, we obtain  
\begin{equation}
\label{eq:Lchi}
  \Delta(\frac{\chi_a}{r})=-4\pi \delta_i+\hat \chi,
\end{equation}
where $\hat \chi$ is a smooth function on $S$ with support in 
$B_a \setminus B_{a/2}$. The ansatz $\theta_0=\chi_a/r +g$ translates the
original equation  into an equation for $g$ 
\[
L_h g = -\hat L_h(\frac{\chi_a}{r}) - \hat \chi.
\]
A direct calculation shows that $\hat L_h(\chi_ar^{-1}) \in L^q(S)$,
$q<3$. By lemma \ref{existenceL} there exists a unique solution 
$g\in W^{2,q}(S)$ to this equation which by the imbedding theorem is in
$C^\alpha(S)$. 

To show that $\theta_0$ is strictly positive, we
observe that it is positive near $i$ (because $r^{-1}$ is
positive and $g$ is bounded) and apply the strong maximum principle
to $-\theta_0$. 
$\blacksquare$\\

We use the conformal covariance of the equation to strengthen the
result on the differentiability of the function $g$. 
Consider a conformal factor 
\begin{equation}
\label{eq:cf}
\omega_0 = e^{f_0} 
\text{ with } f_0 \in C^{\infty}(S) \text{ such that } 
f_0 = \frac{1}{2}\,x^j x^k\,L_{jk}(i)
\text{ on } \quad B_a,
\end{equation}
where we use the normal coordinates $x^k$ and the value of the
tensor
\begin{equation}
\label{eq:Lab}
L_{ab}\equiv R_{ab} -\frac{1}{4}R h_{ab}, 
\end{equation}
at $i$. Then the Ricci tensor of the metric 
\begin{equation}
 \label{eq:h00}
h'_{ab}=\omega_0^4 h_{ab}
\end{equation}
vanishes
at the point $i$ and, since we are in three dimensions, the Riemann
tensor vanishes there too. Hence the connection and metric
coefficients  satisfy in the coordinates $x^k$
\begin{equation}
\label{eq:vR}
\Gamma'_i\,^j\,_k = O(r^2), \quad h'_{ij}=\delta_{ij} + O(r^3).
\end{equation}

\begin{corollary}
\label{impGreen}
The function $g$ found in lemma \ref{existenceGreen} is in
$C^{1,\alpha}(S)$, $0<\alpha<1$.
\end{corollary}

\noindent
{\bf Proof:}
With $\omega_0=e^{f_0}$ and $h'_{ab}$ as above, we note that
\begin{equation}
  \label{eq:yamabere}
  L_{h'}(\theta'_0)=\omega_0^{-5}L_h(\theta_0),
\end{equation}
where 
\begin{equation}
  \label{eq:theta'}
\theta'_0= \omega_0^{-1} \theta_0.
\end{equation}
We apply now the argument of the proof of lemma \ref{existenceGreen} to
the function $\theta'_0$. Since we have by equation (\ref{eq:vR}) 
that $\hat L_{h'}(\chi_ar^{-1}) \in L^\infty(S)$, it follows that
$\theta'_0=\chi_a/r+g'$, where $g'\in C^{1,\alpha}(S)$. We use  
equation (\ref{eq:theta'}), the fact that $\omega_0=1+O(r^2)$, and lemma
\ref{r-1} to obtain the desired result.
$\blacksquare$\\
 
We note that the function 
\begin{equation} 
\label{eq:t-1}
\theta_0^{-1} = \frac{r}{\chi_a+rg},
\end{equation}
is in $C^\alpha (S) $, it is non-negative and vanishes only at $i$.
To obtain $\theta$, we write $\theta = \theta_0 + u$ and solve on $S$
the following equation for $u$
\begin{equation} 
\label{gouu}
L_h u
=-\frac{1}{8} \theta_0^{-7} \Psi_{ab}\Psi^{ab} (1+\theta_0^{-1}u)^{-7}.
\end{equation}

\begin{theorem} 
\label{Beig}
Assume that $h_{ab} \in W^{4,p}(S)$ with $p>3/2$, that $R >0$ on $S$,
and that $\theta_0^{-7}\Psi_{ab}\Psi^{ab}\in L^q(S)$, $q\geq 2$. 
Then there exists a unique non-negative solution $u\in W^{2,q}(S)$  
of equation (\ref{gouu}). We have $u > 0$ on $S$ unless 
$\Psi_{ab}\Psi^{ab} = 0 \in L^q(S)$.
\end{theorem}

We note that our assumptions on $\Psi_{ab}$ impose rather mild
restrictions, which are, in particular, compatible with the fall off
requirement (\ref{Psii}). By the imbedding theorem \ref{imbedding} we
will have $u\in C^\alpha(S)$, $\alpha=2-3/q$, for $q>3$; and $u\in
C^{1,\alpha} (S)$, for $q>3$.

\noindent
{\bf Proof:}
The proof is similar to that given in \cite{Beig}, with the difference
that we impose weaker smoothness requirements. Making use of lemma
\ref{existenceL}, we define a non-linear operator 
$T: B \rightarrow C^0(S)$, with a subset $B$ of $C^0(S)$ which will be
specified below, by setting  
\[
T(u)=L_h^{-1}f(x,u),
\]
where
\begin{equation} \label{eq:fexis}
f(x,u)=-\frac{1}{8}\,\theta_0^{-7}\,\Psi_{ab}\Psi^{ab}\,g(x,u)
\end{equation}
with
\[
g(x,u)=(1+\theta_0^{-1}u)^{-7}.
\]
In the following we will suppress the dependence of $f$ and $g$ on $x$. 
Let $\psi \in W^{2, q}(S) \subset C^{\alpha}(S)$ be the function
satisfying $\psi = T(0)$ and set 
$B = \{u\in C^0(S):\, 0\leq u \leq \psi \}$, which is clearly a closed,
convex subset of the Banach space $C^0(S)$.  

We want to use the Schauder theorem to show the existence of a point $u
\in B$ satisfying $u = T(u)$. This will be the solution to our equation.

We show that $T$ is continuous. Observing the properties of
$\theta_0^{-1}$ noted above, we see that $g$ defines a continuous map
$g: B \rightarrow L^2$. Using the Cauchy-Schwarz inequality, we get  
\[
||f(u_1)-f(u_2)||_{L^2}\leq ||\frac{1}{8} \theta_0^{-7} 
\Psi_{ab}\Psi^{ab}||_{L^2}||g(u_1)-g(u_2)||_{L^2}.
\]
By theorems \ref{existenceL}, \ref{imbedding} we know that the  map
$L^2(S) \rightarrow W^{2,2}(S) \rightarrow C^0(S)$, where the first arrow
denotes the map $L_h^{-1}$ and the second arrow the natural injection,
is continuous. Together these observations give the desired result. 

We show that $T$ maps $B$ into itself. If $u \geq 0$ we have
$f \leq 0$ whence, by lemma \ref{existenceL}, 
$T(u) = L^{- 1}_h(f(u)) \ge 0$. If $u_1 \geq u_2$ it follows 
that $f(u_1) \geq f(u_2)$ whence, again by lemma \ref{existenceL},
$T(u_2) - T(u_1) = L_h^{- 1}(f(u_2) - f(u_1)) \ge 0$. We conclude from
this that for $u \in B$ we have $0 \le T(u) \le T(0) = \psi$.

Finally, $T(B)$ is precompact because $W^{2,2}(S)$ is compactly embedded
in $C^0(S)$ by theorem \ref{Rellich}. Thus the hypotheses of theorem
(\ref{Schauder}) are satisfied and there exists a fix point $u$ of $T$ in
$B$. By its construction we have $u \in L^q(S)$ whence, by elliptic
regularity, $u \in W^{2,q}(S)$.

To show its uniqueness, assume that $u_1$ and $u_2$ are solutions to
(\ref{gouu}). Observing the special structure of $g$ and the identity 
\[
a^{-7} - c^{-7} = (c - a)\,\sum_{j = 0}^6 a^{j - 7} c^{-1 -j},
\]
which holds for positive numbers $a$ and $c$, we find that we can
write $L_h(u_1 - u_2) = c\,(u_1 - u_2)$ with some function $c \ge 0$.
The maximum principle thus allows us to conclude that $u_1 = u_2$. 
The last statement about the positivity of $u$ follows from lemma
\ref{existenceL}, (ii).  
$\blacksquare$

\subsection{Asymptotic expansions near $i$ of solutions to the
Lichnerowicz equation}

The aim of this section is to introduce of the functional spaces 
$E^{m,\alpha}$, to point out the simple consequences listed in Lemma
\ref{comp}, and to proof Theorem \ref{EL}. These are the tools needed to
prove Theorem \ref{T1}.

Let $m\in \mathbb{N}_0$, and denote by  $\mathcal{P}_m$ the space of
homogenous polynomials of degree $m$ in the variables $x^j$. The
element of $\mathcal{P}_m$ are of the form 
$\sum_{|\beta|=m}C_\beta\,x^\beta$ with constant coefficients
$C_\beta$. Note that $r^2$ is in $\mathcal{P}_2$ but $r$ is not in
$\mathcal{P}_1$. We denote by $\mathcal{H}_m$ the set of homogeneous 
harmonic polynomials of degree $m$, i.e the set of $p\in \mathcal{P}_m$ 
such that $\Delta p =0$. For $s\in \mathbb{Z}$, we define the vector 
space  $r^s\mathcal{P}_m$ as the set of functions of the form $r^s  p$
with $p \in \mathcal{P}_m$.

\begin{lemma} 
\label{lE_M}
Assume $s\in \mathbb{Z}$.  The Laplacian defines a bijective linear
map 
\[
\Delta : r^s \mathcal{P}_m \rightarrow r^{s-2} \mathcal{P}_m, 
\]
in either of the following cases

(i) $s>0$

(ii) $s<0$, $|s|$ is odd and  $m+s\geq 0$.
\end{lemma}

Note that the assumptions on $m$ and $s$ imply that the function 
$\Delta(r^s\,p_m) \in C^{\infty}(\mathbb{R}^3 \setminus \{ 0 \})$
defines a function in $L^1_{loc}(\mathbb{R}^3)$ which represents 
$\Delta(r^s\,p_m)$ in the distributional sense.

\noindent
{\bf Proof:} Since $\Delta$ maps $\mathcal{P}_m$ into $\mathcal{P}_{m-2}$,
we find from 
\begin{equation} 
\label{DP}
\Delta (r^s p)=r^{s-2} \left(s(s+1+2m) p +r^2 \Delta  p \right),
\end{equation}
that $\Delta (r^s p) \in r^{s-2} \mathcal{P}_m$. 

We show now that the map is bijective for certain values of $s$ and $m$.  
Because $r^s \mathcal{P}_m$ and $r^{s-2}  \mathcal{P}_m$ have the
same finite dimension, we need only show that the kernel is trivial
for some $s$ and $m$. The vector space $\mathcal{P}_m$ can be written as
a direct sum 
\begin{equation} 
\label{HH}
\mathcal{P}_m=\mathcal{H}_m \oplus r^2  
\mathcal{H}_{m-2}\oplus r^4  \mathcal{H}_{m-4} \cdots ,
\end{equation}
(cf. \cite{Folland}). 
If $\Delta(r^s  p)=0$, we get from (\ref{HH}) that
\[
0 = \Delta(r^s p) = \sum_{0\leq k \leq m/2} \Delta(r^{s+2k} h_{m-2k}),
\]
with  $ h_{m-2k}\in H_{m-2k}$. Applying (\ref{DP}), we obtain 
\[
0 = \sum_{0\leq k \leq m/2} r^{2k} (s+2k)(s+1+2(m-k)) h_{m-2k},
\]
which allows us to conclude by (\ref{HH}) that 
\[
(s+2k)(s+1+2(m-k)) h_{m-2k}=0.
\]
Since by our assumptions $(s+2k)(s+1+2(m-k)) \neq 0$, it follows that
the polynomials $h_{m-2k}$ vanish, whence $r^s p = 0$. 
$\blacksquare$\\

We will need the following technical Lemma regarding H\"older functions: 
\begin{lemma} 
\label{r-1}
Suppose $m \in \mathbb{N}$, $0 < \alpha < 1$,
$f \in C^{m,\alpha}(B_a)$, and $T_m$ denotes
the Taylor polynomial of order $m$ of $f$.  
Then $f_R \equiv f - T_{m}$ is in $C^{m,\alpha}(B_a)$ and satisfies,
if $|\beta| \le m$, 
\[
\partial^\beta f_R = O(r^{m-|\beta|+\alpha})
\quad\text{as}\quad r \rightarrow 0.
\]
Moreover, let $s$ be an integer such that $s\leq 1$ and  
$m+s-1\geq 0$. Then $f_R$ satisfies: 

(i)
$r^{s-2}f_R\in W^{m+s-1,p}(B_\epsilon)$, for $p<3/(1-\alpha)$,  
$0<\epsilon<a$.

(ii)
 If $m+s-1 \geq 1$ then $r^{s-2}f_R\in C^{m+s-2,\alpha}(B_\epsilon)$.  

(iii)
$rf_R \in C^{m,\alpha}(B_\epsilon)$.
\end{lemma}

\noindent
{\bf Proof:}
The relation
\begin{equation} \label{Of}
|\partial^\beta f_R|\leq C|x|^{m-|\beta|+\alpha},\,\,\,\,\,\,\,
x\in \bar B_\epsilon, 
\end{equation}
is a consequence of Lemma \ref{holderrest}.   

(i) We have
\[
\partial^\beta (r^{s-2}f_R) 
= \sum_{\beta' + \gamma' = \beta }C_{\beta'} 
\partial^{\beta'} f_R\,\partial^{\gamma'} (r^{s-2}),
\]
with certain constants $C_{\beta'}$, and the derivatives of $r^{s-2}$ are
bounded for $x \in  \bar B_\epsilon$ by 
\[
|\partial^{\gamma'} r^{s-2}| \leq  C r^{s-2-|\gamma'|}.
\]
Observing (\ref{Of}), we obtain
\[
|\partial^\beta (r^{s-2}f_R)|\leq C{r^{m-|\beta|+s-2+\alpha}},
\]
for $x \in \bar B_\epsilon$, whence, by our hypothesis $m+s-1\geq 0$,
\begin{equation}
\label{eq:s-2fR}
|\partial^\beta (r^{s-2}f_R)|\leq C{r^{-1+\alpha}},
\end{equation}
for $|\beta|\leq m+s-1$. Using that $r^{-1+\alpha}$ is in
$L^p(B_\epsilon)$ for
$p<3/(1-\alpha)$, we conclude that  $\partial^\beta (r^{s-2}f) \in
L^p(B_\epsilon)$ whence $r^{s-2}f_R \in W^{m+s-1,p}(B_\epsilon)$. 

(ii) From the relation above and theorem \ref{imbedding} we conclude for 
$m+s-1 \geq 1$ that $r^{s-2}f_R \in C^{m+s-2, \alpha '}(B_\epsilon)$ with 
$\alpha'=1-3/p < \alpha$. To show that $\alpha '$ can in fact be
chosen equal to $\alpha$, we use the sharp result of theorem
\ref{Morrey}.  Set $g= \partial ^{\beta '}(r^{s-2}f_R)$ for some $\beta
'\leq m+s-2$. Let $z$ be an arbitrary point of $B_\epsilon$ and $B_R(z)$
be a ball with center $z$ and radius $R$ such that $B_R \subset B_a$.   
Using  the inequality (\ref{eq:s-2fR}), we obtain
\[
\int_{B_R(z)} |\partial g| d\mu \leq C\int_{B_R(z)}  r^{\alpha -1} d\mu.
\leq C\int_{B_R(0)} r^{\alpha -1} d\mu = C'R^{2+\alpha}
\]
(cf. \cite{Gilbarg} p. 159 for the second estimate). Applying now Lemma
\ref{Morrey}, we conclude that $g\in C^\alpha(B_\epsilon)$, whence
$r^{s-2}f_R \in C^{m+s-2, \alpha}(B_\epsilon)$.

(iii)
 We have 
\[
\partial^\beta (rf_R)=r \partial^\beta f_R +f_1,
\]
where, with certain constants $C_{\beta,\beta'}$, 
\[
f_1=\sum_{\beta \neq \beta' + \gamma' \le \beta} 
C_{\beta,\beta'} \partial^{\beta'}f_R \partial^{\gamma'}r.
\]
Note that $r\partial^\beta f \in C^\alpha(B_a)$, since $r$ is Lipschitz 
continuous.

Using the bound
\[
|\partial^{\gamma'} r|\leq Cr^{-|\gamma'|+1},
\]
and the bound (\ref{Of}) 
we obtain 
\[
|\partial f_1|\leq C r^\alpha.
\]
Thus $f_1 \in W^{1,p}$ for all $p$, whence, by theorem \ref{imbedding},
$f_1 \in C^\alpha$ for $\alpha<1$. 
$\blacksquare$\\

The following function spaces will be important for us. 
\begin{definition} 
\label{dEm}
For $m\in \mathbb{N}_0$ and $0<\alpha <1$, we define the space 
$E^{m,\alpha}(B_a)$ as the set
$E^{m,\alpha}(B_a)= \{ f=f_1+rf_2  \, : \,  f_1,f_2  \in C^{m,\alpha}(B_a) \}$.
Furthermore we set 
$E^{\infty}(B_a)= \{ f=f_1+rf_2  \, : \,  f_1,f_2  \in C^{\infty}(B_a) \}$.
\end{definition}

Note that the decompositions above are not unique. If
$f=f_1+rf_2$, $f_1,f_2  \in C^{m,\alpha}(B_a)$ then also
$f=f_1 + rf_R +r(f_2 - f_R)$ with 
$f_1 + rf_R, f_2 - f_R \in C^{m,\alpha}(B_a)$ if $f_R$ is given as in 
lemma \ref{r-1}. Obviously, 
$E^{\infty}(B_a) \subset E^{m,\alpha}(B_a)$ for all $m\in \mathbb{N}_0$.
The converse is not quite immediate. 
\begin{lemma}
\label{Einfty}
If $f\in E^{m,\alpha}(B_a)$ for all $m\in \mathbb{N}_0$, then 
$f\in E^\infty(B_a)$.
\end{lemma}

\noindent
{\bf Proof:}
Assume that $f\in E^{m,\alpha}(B_a)$ for all $m$. Take an arbitrary $m$
and write $f=f_1+rf_2$ with $f_1,f_2\in C^{m,\alpha}(B_a)$. To obtain a
unique representation, we write $f_1$ and $f_2$ as the sum of their Taylor
polynomials of order $m$ and their remainders, 
\begin{equation}
\label{eq:splitf1f2}
f_1=\sum_{j=0}^m p^1_j+ f^1_R, \quad
f_2=\sum_{j=0}^m p^2_j+ f^2_R,
\end{equation}
with $p^1_j, p^2_j \in\mathcal{P}_j$ 
and $f^1_R, f^2_R = O(r^{m+\alpha})$. From this we get the representation
\begin{equation}
\label{eq:splitf}
f= (\sum_{j=0}^mp^1_j+r\sum_{j=0}^{m-1} p^2_j) +f_R, 
\end{equation} 
where $f_R \equiv f^1_R+r(f^2_R+p^2_m) \in C^{m,\alpha}(B_a)$ and
$f_R = O(r^{m+\alpha})$ by lemma \ref{r-1}. This decomposition is unique:
if we had $f = 0$, the fast fall-off of $f_R$ at the origin would imply
that the term in brackets, whence also each of the polynomials and $f_R$,
must vanish.

Since $m$ was arbitrary, we conclude that the function $f$ determines a
unique sequence of polynomials $p^2_j$, $j \in \mathbb{N}_0$ as above. 
By Borel's theorem  (cf. \cite{Dieudone69}) there exists a function 
$v_2 \in C^{\infty}(B_a)$ (not unique) such that   
\begin{equation}
\label{2Borel}
v_2 - \sum_{j=0}^{m} p^2_j = O(r^{m + 1}),\,\, m \in \mathbb{N}_0. 
\end{equation}
We show that the function $v_1 \equiv f-rv_2$ is $C^{m-1}(B_a)$ for
arbitrary $m$, i.e. $v_1 \in C^{\infty}(B_a)$. Using (\ref{eq:splitf}), 
we obtain
\begin{equation}
\label{eq:v2Borel}
v_1= \left(\sum_{j=0}^mp^1_j +f_R\right)
+ r\left(\sum_{j=0}^{m-1} p^2_j-v_2\right).
\end{equation}
The first term is in $C^{m,\alpha}(B_a)$ by the observations above,
the second term is in $C^{m,\alpha}(B_a)$ by (\ref{2Borel}) and 
lemma \ref{r-1}. 
$\blacksquare$\\

While we cannot directly apply elliptic regularity
results to these spaces, they are nevertheless appropriate for our purposes.
This follows from the following observation, which will be extended to more
general elliptic equations and more general smoothness assumptions in
theorem \ref{EL} and in appendix \ref{additional}.

If $u$ is a solution to the Poisson equation
\[
\Delta u = \frac{f}{r},
\]
with $f\in E^{m,\alpha}(B_a)$, $m\geq 1$, then 
$u \in E^{m+1,\alpha}(B_a)$. This can be seen as follows. If we write 
$f = f_1 + r\,f_2 \in E^{m,\alpha}(B_a)$ in the form   
\[
\frac{f}{r}=\frac{T_m}{r} + f_R,
\] 
where $T_m$ is the Taylor polynomial of order $m$ of $f_1$, the
remainder $f_R$ is seen to be in $C^{m-1, \alpha}(B_a)$ by Lemma
\ref{r-1}. 

By Lemma  \ref{lE_M}, there exist a 
polynomial $\hat T_m$ such that
\[
\Delta (r\hat T_m) = \frac{T_m}{r}.
\]
Then $u_R \equiv u - r\hat T_m$ satisfies $\Delta u_R = f_R$
and theorem \ref{SchauderEstimate} implies that  
$u_R \in C^{m+1, \alpha}(B_a)$, whence $u \in E^{m+1, \alpha}(B_a)$.

To generalize these arguments to equations with non-constant coefficients
and to non-linear equations we note the following observation.

\begin{lemma} \label{comp}
For $f, g \in E^{m,\alpha}(B_a)$ we have

(i) $f+g\in E^{m,\alpha}(B_a)$ 

(ii)  $fg  \in  E^{m,\alpha}(B_a) $

(iii) If  $f \neq 0$ in $B_a$, then $1/f \in  E^{m,\alpha}(B_a)$.

Analogous results hold for functions in $E^{\infty}(B_a)$.  
\end{lemma}

\noindent
{\bf Proof:}
The first two assertions are obvious, for (iii) we need only consider
a small ball $B_\epsilon$ centered at the origin because $r$ is smooth
elsewhere.  If $f=f_1 +rf_2$, $f_1,f_2 \in C^{m,\alpha}(B_a)$, we have
$1/f=v_1 +rv_2$ with
\[
v_1=\frac{f_1}{(f_1)^2-r^2(f_2)^2},\,\,\,\,\,\,\,
v_2=\frac{- f_2}{(f_1)^2-r^2(f_2)^2}.
\]
These functions are in $C^{m,\alpha}(B_\epsilon)$ for sufficiently small 
$\epsilon > 0$ because our assumptions imply that $f_1(0)\neq 0$.
The $E^{\infty}(B_a)$ case is similar.
$\blacksquare$\\

We consider now a general linear elliptic differential operator $L$ of
second order  
\begin{equation}
  \label{eq:L2o}
  L=a^{ij} \partial_i  \partial_j +b^i  \partial_i +c.
\end{equation}
It will be assumed in this section that
\begin{equation}
  \label{eq:smoothc}
  a^{ij},\, b^i, \, c \in C^{\infty}(\bar B_a).
\end{equation}

We express the operator in normal geodesic coordinates
$x^i$ with respect to $a^{ij}$, centered at the origin of $B_a$, such that
\[
a^{ij}(x)= \delta^{ij} + \hat a^{ij},
\]
with
\begin{equation} \label{Oaij}
\hat a^{ij}=O(r^2),
\end{equation}
and 
\begin{equation} \label{xxaij}
 x_j \hat a^{ij}=0,\,\,\,\,\,x \in B_a.
\end{equation}

For the differential operator $\hat L$, given by
\[
\hat Lu = 
\hat a^{ij}(x) \partial_i \partial_j u + b^i(x) \partial_i u + c(x) u,
\]
we find
\begin{lemma} 
\label{Lhat}
Suppose $p \in \mathcal{P}_m$. Then the function $U$ defined by 
$\hat L(r^s p)=r^{s-2}U$ is $C^\infty$ and satisfies $U=O(r^{m+1})$. 
If in addition $b^i=O(r)$ (as in the case of the Yamabe
operator $L_h$), then $U=O(r^{m+2})$.
\end{lemma}

\noindent
{\bf Proof:}
A direct calculation, observing (\ref{xxaij}), gives   
\[
U = \hat a^{ij}(s \delta_{ij}p+r^2 \partial_j \partial_i p)  
+b^i(sx_ip+r^2\partial_i p)+cr^2p,
\]
which entails our result. 
$\blacksquare$\\

In the following we shall use the splitting $L=\Delta +\hat L$,
where $\Delta$ is the flat Laplacian in the normal coordinates $x^i$.

\begin{theorem} 
\label{EL}
Let $u \in W^{2,p}_{loc}(B_a)$ be a solution of
\[
Lu=r^{s-2}f,
\]
where $L$ is given by (\ref{eq:L2o}) with (\ref{eq:smoothc}) and $s \in \mathbb{Z}$,
$p>1$. 

(i) Assume $s=1$ and $f\in E^{m,\alpha}(B_a)$. 
Then $u \in E^{1,\alpha '}(B_a)$,
$0 < \alpha' < \alpha$, if $m=0$ and
$u\in E^{m+1,\alpha}(B_a)$ if $m\geq 1$.  If  $f \in E^\infty(B_a)$, then  $u \in E^\infty(B_a)$.

(ii) If $s<0$, $|s|$ is odd, $f\in C^{m,\alpha}(B_a)$ with $m+s-1\geq 0$,
and $f=O(r^{s_0})$, with $s_0+s\geq 0$; then $u$ has the form
\[
u=r^s\sum_{k=s_0}^m u_k +u_R,
\]
where $u_k \in \mathcal{P}_m$, $Lu_R =O(r^{m+\alpha+s-2})$, and
$u_R \in C^{1,\alpha'}(B_a)$, $0 < \alpha' < \alpha$, if $m+s-1 = 0$,
$u_R \in C^{m+s,\alpha}(B_a)$ if $m+s-1 \geq 1$. 

If $f \in C^{\infty}(B_a)$, then    
then $u$ can be written in the form $u=r^sv_1+v_2$ with 
$v_1, v_2 \in C^\infty(B_a)$, $v_1=O(r^{s_0})$,  and  
\[
L(r^sv_1)=r^{s-2}f +\theta_R, \quad L(v_2)= -\,\theta_R,
\]
where $\theta_R \in C^\infty(B_a)$ and all its derivatives vanish at
the origin.
\end{theorem}

\noindent
{\bf Proof:}
In both cases we write $f=T_{m}+f_R$ with a polynomial $T_m$ of order $m$,
\[
T_{m}=\sum_{k=m_0}^{m} t_k,\,\,\,\,\,\,\,
t_k \in \mathcal{P}_m.
\]
Case (i): $m_0=0$  and  $f=f_1+rf_2$, where $f_1, f_2$ are in
 $ C^{m,\alpha}(B_a)$. We define $T_m$ to be the Taylor polynomial of
 order $m$ of $f_1$.\\
Case (ii): $m_0=s_0$ and we define $T_m$ to be the Taylor polynomial of
order $m$ of $f$. 

We show that $u$ has in both cases the form
\begin{equation} 
\label{su}
u=r^s \sum_{k=m_0}^{m} u_k +u_R ,
\end{equation}
with $u_k \in \mathcal{P}_k$ and $u_R\in C^{m+s,\alpha}(B_a)$. For this
purpose lemma \ref{lE_M} will be used to determine the polynomials $u_k$
in terms of $t_k$ by a recurrence relation. The differentiability of
$u_R$ follows then from lemma \ref{r-1}, elliptic regularity, and
the elliptic equation satisfied by $u_R$.

The recurrence relation is defined by 
\begin{equation} 
\label{recur}
\Delta(r^s u_{m_0})=r^{s-2}t_{m_0},\,\,\,\,\,\,\,
\Delta(r^su_k)=r^{s-2}(t_k-U^{(k)}_k),\,\,\,\,\,\,\,
k = m_0 + 1, \ldots, m,
\end{equation}
where, given $u_{m_0}, \ldots, u_{k - 1}$, we define $U^{(k)}_k$ as follows.   
By Lemma (\ref{Lhat}) the functions
\begin{equation} 
\label{U_k}
U^{(k)} = r^{-s+2}\,\hat L (r^s\sum_{j=m_0}^{k-1} u_j), 
\end{equation}
which will be defined successively for $k = m_0 + 1, \ldots, m + 1$, 
are $C^\infty$ and $U^{(k)} = O(r^{m_0+1})$. Thus we can write by lemma
(\ref{holderrest}) 
\[
U^{(k)}=\sum_{j=m_0+1}^m  U^{(k)}_j +U^{(k)}_R,
\]
where $U^{(k)}_R=O(r^{m+\alpha})$ and $U^{(k)}_j \in \mathcal{P}_j$ denotes
the homogenous polynomial of order $j$ in the Taylor expansion of $U^{(k)}$.

By Lemma (\ref{lE_M}) the recurrence relation (\ref{recur}) is well defined 
for the cases (i) and (ii). Note that
\begin{equation} 
\label{kj} 
U^{(k')}_j=U^{(k)}_j,\,\,\,m_0 + 1 \leq j \leq k
\quad\mbox{if}\quad k < k' \leq m + 1,
\end{equation}
because we have by Lemma \ref{Lhat}
\[
U^{(k')}-U^{(k)}=r^{-s+2}\hat L (r^s\sum_{j=k}^{k'-1} u_j)=O(r^{k + 1}).
\]

With the definitions above and the identity (\ref{kj}), which allows us to replace
$U^{(k)}_k$ by $U^{(m+1)}_k$, the original equation for $u$ implies for the function
$u_R$ defined by (\ref{su}) the equation 
\begin{equation} 
\label{fuR}
Lu_R=r^{s-2}\left(U^{(m+1)}_R  + f_R \right).
\end{equation}

Case (i): 
We use lemma (\ref{r-1}), (i) to conclude that the right hand side of 
equation (\ref{fuR}) is in $L^p(B_\epsilon)$ if $m=0$ and in 
$C^{m-1,\alpha}(B_\epsilon)$ if $m\geq 1$. 
Now theorems \ref{globalregularity} and \ref{SchauderEstimate} imply 
that $u_R \in C^{1,\alpha'}(B_a)$, $\alpha'< \alpha$, if $m=0$
and $u_R \in C^{m+1,\alpha}(B_a)$ if $m\geq 1$. For the $E^\infty$
case we use lemma \ref{Einfty}.

Case (ii): By our procedure we have 
\[
Lu_R =O(r^{m+\alpha+s-2}).
\]
If $m + s - 1 = 0$ we use part (i) of lemma \ref{r-1} to conclude that the
right hand side of equation (\ref{fuR}) is in $L^p(B_a)$ and theorems
\ref{globalregularity}, \ref{imbedding} to conclude that 
$u_R \in C^{1, \alpha'}(B_a)$. 
If $m + s - 1 \ge 1$ we use part (ii) of lemma \ref{r-1} to conclude
that the right-hand side of equation (\ref{fuR}) is in 
$C^{m+s-2,\alpha}(B_a)$.  Elliptic regularity then implies that
$u_R \in C^{m+s,\alpha}(B_a)$. The $C^\infty$ case follows by an
analogous argument as in the proof of lemma  \ref{Einfty}, since  
the polynomials $u_{m_0}, \ldots, u_m$ obtained for an integer
$m'$ with $m' > m$ coincide with those obtained for $m$, i.e. the 
procedure provides a unique sequence of polynomials $u_k$, 
$k = m_0, \ldots \,\,$
$\blacksquare$\\

More general expansions, which include logarithmic terms, have been 
studied (in a somewhat different setting) in \cite{Meyers}, where results
similar to those given in \ref{EL} have been derived. Definition
\ref{dEm} is taylored to the case in which no logarithmic terms appear
and leads to a considerable simplification of the proofs as well as to a
more concise statement of the results as compared with those given in
\cite{Meyers}.

\begin{corollary} 
\label{Green}
Assume that the hypotheses on $u$ of theorem \ref{EL} are satisfied. Let
$\theta_0$  be a  distributional solution of $L\,\theta_0=-4\pi \delta_i$
in  $B_a$. Then we can write
\begin{equation} 
\label{the}
\theta_0=r^{-1}u_1+u_2,
\end{equation}
with $u_1,u_2 \in C^\infty(B_a)$, $u_1(0)=1$, 
$L(r^{-1}u_1)=-4\pi \delta_i +\theta_R$, where 
$\theta_R \in C^{\infty}(B_a)$ and all its derivatives vanish  at
$i$. In the particular case of the Yamabe operator $L_h$ with respect 
to a smooth metric $h$ we have  $u_1=1+O(r^2)$. 
\end{corollary}

\noindent
{\bf Proof:}
Using that $\Delta (r^{-1})=-4\pi \delta_i$ in $B_a$, 
we obtain for $u = \theta_0 - 1/r$
\begin{equation} \label{equu}
Lu=-\hat L(r^{-1}).
\end{equation}
By lemma \ref{Lhat} we have $\hat L(r^{-1})=r^{-3}U$,
with $U \in C^\infty(B_a)$ and $U=O(r)$. Our assertion now follows from
theorem \ref{EL}. For the last  assertion we use that in the case of
$L_h$  we have $U=O(r^2)$.  
$\blacksquare$\\

We note that the functions $u_1$, $u_2$ are in fact analytic 
and $\theta_R \equiv 0$ if the coefficients of $L$ are analytic in $B_a$
(cf. \cite{Garabedian}). 

\subsection{Proof of Theorem \ref{T1}}

There exists a unique solution $\theta=\theta_0+u$ of the equation
(\ref{Lich}) with $\theta_0$ as lemma \ref{existenceGreen} and $u$ as in
theorem \ref{Beig}. Since the operator $L_h$ satisfies the hypothesis of
corollary \ref{Green} we can write on $B_a$
\begin{equation} 
\label{the0}
\theta_0 =\frac{u_1}{r}+w,
\end{equation}
where $u_1$, $w$ are smooth functions and 
$L_h (r^{-1}u_1)=\theta_R$ on $B_a \setminus {i}$, with $\theta_R$ as
described in corollary \ref{Green}.
 
Given the solution $u = u(x)$, we can read equation (\ref{gouu})  
in $B_a$ as an equation for $u$ 
\begin{equation} 
\label{u}
L_h  u =\frac{f(x)}{r},
\end{equation}
with $f$ considered as a given function of $x$
\[
f(x) =-\frac{r^8 \Psi_{ab}\Psi^{ab}}{8(r\theta_0+ru)^7}.
\]

By hypothesis $r^8\Psi_{ab}\Psi^{ab} \in E^\infty(B_a)$, by
corollary \ref{Green} we have $r\,\theta_0 \in E^\infty(B_a)$, and 
by theorem \ref{Beig} the solution $u$ is in
$C^\alpha(B_a) \subset E^{0,\alpha}(B_a)$. By lemma \ref{comp} we thus
have $f \in E^{0,\alpha}(B_a)$. By theorem \ref{EL} equation
(\ref{u}) implies that $u \in E^{1,\alpha'}(B_a)$, $0<\alpha'<\alpha$
which implies in turn that $f \in E^{1,\alpha'}(B_a)$.  
Repeating the argument, we show inductively that 
$u$, whence $f$ is in $E^{m, \alpha '}(B_a)$ for all $m \ge 0$. 
Lemma \ref{Einfty} now implies that $u \in E^\infty(B_a)$.
$\blacksquare$

\subsection{Solution of the Hamiltonian Constraint with Logarithmic 
terms} 
\label{logs}

The example 
\begin{equation} \label{logg}
\Delta(\log r  h_m)=r^{-2} h_m (2m+1),
\end{equation}
$h_m \in \mathcal{H}_m$, shows that logarithmic terms can occur
in solutions to the Poisson equation even if the source has only terms of
the form $r^sp$ with $p \in \mathcal{P}_m$. This happens in the cases
where the Laplacian does not define a bijection between
$r^{s + 2}\mathcal{P}_m$ and $r^s\mathcal{P}_m$, cases which are
excluded in lemma \ref{lE_M}. We shall use this to show that logarithmic
terms can occur in the solution to the Lichnerowicz equation if the
condition $r^8\Psi_{ab}\Psi^{ab} \in E^\infty (B_a)$ is not satisfied.
Our example will be concerned with initial data with non-vanishing
linear momentum.

We assume that in a small ball $B_a$ centered at $i$
the tensor $\Psi^{ab}$ has the form 
\begin{equation}
\label{eq:pcr}
\Psi^{ab}=\Psi_P^{ab}+\Psi_R^{ab},
\end{equation}
where $\Psi_P^{ab}$ is given in normal coordinates by (\ref{eq:PsiPp}) 
and $\Psi_R^{ab}=O(r^{-3})$ is a tensor field such that
$\Psi^{ab}$ satisfies equation (\ref{diver}). The existence of such
tensors, which satisfy also
\begin{equation}
\label{eq:gcpsi}
\theta_0^{-7} \Psi_{ab}\Psi^{ab}\in L^q(S), \quad q\geq 2,
\end{equation}
and, by lemma \ref{PsiPsi},
\begin{equation}
\label{eq:GG}
r^8\Psi_{ab}\Psi^{ab}=  \psi +r\psi^R \quad\mbox{in}\quad B_a,
\end{equation}
will be shown in section \ref{asymtoticmomentum}. Here the function
$\psi^R$ is in $C^{\alpha}(B_a)$ and $\psi$ is given explicity by
\begin{equation} 
\label{G}
\psi \equiv r^8\Psi_{P\,ij}\Psi_P^{ij}=c+r^{-2} h_2,
\end{equation}
with $c= \frac{15}{16}\,P^2$, $P^2=P^iP_i$, and
\begin{equation}
\label{eq:h2}
h_2=\frac{3}{8} r^2 \left(3(P^i n_i)^2-P^2  \right),
\end{equation}
where, in accordance with the calculations in section
\ref{asymtoticmomentum},
the latin indices in the expressions above are moved with the flat
metric. We note that $h_2 \in  \mathcal{H}_2$ and $\psi$ is not in
continuous. The tensor $\Psi^{ab}$ satisfies condition (\ref{Psii}) and 
the three  constants $P^i$ given by (\ref{eq:PsiPp}) represent the 
momentum of the initial data. Since $\Psi_P^{ab}$ is trace-free and
divergence-free with respect to the flat metric, we could, of course,
choose $h_{ab}$ to be the flat metric and $\Psi_R^{ab}=0$. This would
provide one of the conformally flat initial data sets discussed in
\cite{York}. We are interested in a more general situation.

\begin{lemma}
Let $h_{ab}$ be a smooth metric, and let $\Psi^{ab}$ be given by 
(\ref{eq:pcr}) such that conditions (\ref{eq:gcpsi}) and (\ref{eq:GG})
hold. Then, there exists a unique, positive, solution to the Hamiltonian
constraint (\ref{Lich}). In $B_a$ it has the form 
\begin{multline} 
\label{eq:logexp}
\theta= \frac{w_1}{r} + \frac{1}{2}m 
+ \frac{1}{32}\,r\,\left((9\,(P^i n_i)^2 - 33\,P^2 \right) \\
+ \frac{7}{16}\,m\,r^2\,\left(\frac{5}{4}\,P^2 
+ \frac{3}{5}\,(3\,(P^i n_i)^2-P^2)\,\log r \right) + u_R,
\end{multline}
where the constant $m$ is the total mass of the initial data, $w_1$ is 
a smooth function with $w_1=1 + O(r^2)$, and 
$u_R\in C^{2,\alpha}(B_a)$ with  $u_R(0)=0$.
\end{lemma}

Since $w_1$ is smooth and $u_R$ is in $C^{2,\alpha}(B_a)$, there cannot
occur a cancellation of logarithmic terms. For non-trivial data,
for which $m \neq 0$, the logarithmic term will always appear. In the
case where $h_{ab}$ is flat and $\Psi_R^{ab}=0$ an expansion similar to
(\ref{eq:logexp}) has been calculated in \cite{Gleiser99}. 

\noindent
{\bf Proof:}
The existence and uniqueness of the solution has been shown in section
\ref{hexistence}. To derive (\ref{eq:logexp}) we shall try to calculate
each term of the expansion and to control the remainder as we did in the
proof of theorem \ref{EL}. However, lemma \ref{lE_M} will not suffice
here, we will have to use equation (\ref{logg}).

By  Corollary \ref{Green} we have 
\[
\theta_0=\frac{w_1}{r} +w,
\]
with $w, w_1 \in C^{\infty}(B_a)$ and $w_1=1 + O(r^2)$. By theorem
\ref{Beig} the unique solution $u$ of equation (\ref{u}) is in
$C^\alpha(B_a)$. Equation (\ref{u}) has the form
\[
L_h u = (\frac{\psi}{r}+\psi^R)\,f(x,u) \quad\mbox{with}\quad
f(x,u)=-\frac{1}{(w_1+r(u+w))^7}.
\]
By $m \equiv 2\,(u(0)+w(0))$ is given the mass of the initial data. 
Since $u \in C^\alpha(B_a)$, we find
\begin{equation}
\label{eq:sf}
f=-1+\frac{7}{2}\,m\,r+f_R,\,\,\,\,\,\,f_R=O(r^{1+\alpha}).  
\end{equation}
If we set 
\begin{equation}
\label{eq:logu1}
u_1= -\frac{c}{2} r+\frac{1}{4r}h_2 +
\frac{7}{2}\,m\,\left(\frac{c}{6}\,r^2 
+ \frac{1}{5}\,\log r\,h_2\right),
\end{equation}
we find from (\ref{DP}), (\ref{logg}) that  
\begin{equation}
\label{eq:Dlog}
\Delta u_1= \frac{\psi}{r} (-1+ \frac{7}{2}\,m\,r),
\end{equation}
and that $v = u - u_1$ satisfies
\begin{equation}
\label{eq:logv}
 L_h v=\psi^R(-1+ \frac{7}{2}\,m\,r)
-\hat L_h u_1+\frac{\psi f_R}{r}.
\end{equation}

We shall show that this equation implies that the function $v$ is in
$C^{2,\alpha}(S)$. Since $\psi^R\in C^\alpha(B_a)$, the first term on
the right hand side of (\ref{eq:logv}) is in $C^\alpha(B_a)$. By a direct
calculation (using that the coefficient $b^i$ of $\hat L_h$ is $O(r)$)  
we find that $\hat L_h u_1 \in W^{2,p}(B_a)$, $p<3$, which implies by
the Sobolev imbedding theorem that $\hat L_h u_1 \in C^\alpha(B_a)$. 
The third term on the right-hand side of (\ref{eq:logv}) is more
delicate, because it depends on the solution $v$. From theorem
\ref{Beig} and equation (\ref{eq:logu1}) we find that $v \in C^\alpha(B_a) \cap
W^{2,p}(B_a)$, where, due to (\ref{eq:logu1}), we need to
assume $p < 3$. However, since $f_R=O(r^{1+\alpha})$ and $\psi$ is
bounded, the  function $\psi f_R/r$ is bounded and the right hand side of
(\ref{eq:logv}) is in $L^\infty(B_a)$. Thus
theorem \ref{globalregularity} implies that 
$v\in W^{2,p}(B_a)$ for $p > 3$, whence $v\in C^{1,\alpha}(B_a)$ by the
Sobolev imbedding theorem. It follows that we can differentiate $f_R$,
considered as a function of $x$, to find that $\partial_i\,f_R=O(r)$.
It follows that $\psi f_R/r$ is in $W^{1,p}(B_a)$, $p > 3$, and thus
in $C^\alpha(B_a)$. Since the right hand side of (\ref{eq:logv}) is in
$C^\alpha(B_a)$, it follows that $v\in C^{2,\alpha}(B_a)$.
$\blacksquare$

\section{The Momentum Constraint}

\subsection{The Momentum Constraint on Euclidean Space} 
\label{flatmomentum}

In the following we shall give an explicit constructing of the
smooth solutions to the equation $\partial_a\Psi^{ab}=0$ on 
3-dimensional Euclidean space $\mathbb{E}^3$ or open subsets of it.
Another method to obtain such solutions has been described in
\cite{Beig96'}, multipole expansions of such tensors have been studied in
\cite{Beig96}. 

Let $i$ be a point of $\mathbb{E}^3$ and $x^k$ a Cartesian coordinate
system with origin $i$ such that in these coordinates the metric of
$\mathbb{E}^3$, denoted by $\delta_{ab}$, is given by the standard
form $\delta_{kl}$. We denote by $n^a$ the vector field on 
$\mathbb{E}^3 \setminus \{i\}$ which is given in these coordinates by
$x^k/|x|$.

Denote by $m_a$ and its complex conjugate $\bar{m}_a$ complex vector
fields, defined on $\mathbb{E}^3$ outside a lower dimensional subset and
independent of $r = |x|$, such that 
\begin{equation} 
\label{nm}
m_a m^a=\bar{m}_a\bar{m}^a=n_am^a=n_a\bar{m}^a=0,\,\,\,\,\,\,
m_a \bar{m}^a=1.
\end{equation}
There remains the freedom to perform rotations
$m_a \rightarrow e^{if} m_a$
with functions $f$ which are independent of $r$. 

The metric has the expansion
\[
\delta_{ab}=n_an_b+\bar m_a m_b+ m_a\bar m_b,
\]
while an arbitrary symmetric, trace-free tensor $\Psi_{ab}$ can be
expended in the form 
\begin{multline} 
\label{Psit}
r^{3}\Psi_{ab}= \xi(3n_a n_b -\delta_{ab})+\sqrt{2}\eta_1 
n_{(a} \bar m_{b)}+ \sqrt{2}\bar \eta_1 n_{(a} m_{b)}+ \\
\bar \mu_2 m_a 
m_b+\mu_2 \bar m_a \bar  m_b,
\end{multline}
with
\[
\xi = \frac{1}{2}r^3 \Psi_{ab} n^a n^b, \quad
\eta_1 = \sqrt{2} r^3 \Psi_{ab} n^a m^b, \quad
\mu_2 = r^3 \Psi_{ab} m^a m^b.
\]
Since  $\Psi_{ab}$ is real, the function $\xi$ is real while $\eta_1$, 
$\mu_2$ are complex functions of spin weight 1 and 2 respectively.

Using in the equation
\begin{equation} 
\label{divp}  
\partial_a \Psi^{ab}=0, 
\end{equation}
the expansion (\ref{Psit}) and contracting suitably with $n^a$ and 
$m^a$, we obtain the following representation of
(\ref{divp})   
\begin{equation} 
\label{n1}
4r\partial_r \xi + \bar \eth \eta_1 +\eth \bar \eta_1=0,
\end{equation}
\begin{equation} 
\label{q1}
r\partial_r \eta_1 + \bar \eth \mu_2 - \eth \xi =0.
\end{equation}
Here $\partial_r$ denotes the radial derivative and $\eth$ the edth
operator of the unit two-sphere (cf. \cite{Penrose} for
definition and properties). By our assumptions the differential operator
$\eth$ commutes with $\partial_r$. 

Let ${}_sY_{lm}$ denote the spin weighted spherical harmonics, which
coincide with the standard spherical harmonics $Y_{lm}$ for $s=0$. The
${}_sY_{lm}$ are  eigenfunctions of the operator $\bar \eth \eth$ for
each spin weight $s$.
More generally, we have
\begin{equation}
\label{eq:betheth2}
\bar \eth^p \eth^p ({}_sY_{lm})=(-1)^p \frac{(l-s)!}{(l-s-p)!}\frac{(l+s+p)!}{(l+s)!} {}_sY_{lm}.
\end{equation}

If $\mu_s$ denotes a smooth function on the two-sphere of integral
spin weight $s$, there exists a function $\mu$ of spin weight zero such
that 
$\eta_s=\eth^s \eta$. We set
$\eta^R=\textnormal{Re}(\eta)$ and
$\eta^I=i\,\textnormal{Im}(\eta)$, such that 
$\eta= \eta^R + \eta^I$, and define 
\[
\eta_s^R=\eth^s \eta^R, \,\,\, \eta_s^I=\eth^s \eta^I,
\]
such that $\eta_s= \eta_s^R +\eta_s^I$. It holds
\[
\overline{\eth^s\eta^R} = \bar\eth^s \eta^R, \,\,\, 
\overline{\eth^s\eta^I} = - \bar\eth^s \eta^I.
\]

Using these decompositions now for $\eta_1$ and $\mu_2$, we obtain 
equation (\ref{n1}) in the form 
\begin{equation} 
\label{e1}
2r\partial_r \xi =- \bar  \eth  \eth \eta^R.
\end{equation}
Applying $\bar \eth$ to both sides of equation (\ref{q1}) and
decomposing into real and imaginary part yields 
\begin{equation} 
\label{e2}
r\partial_r \bar \eth \eth \eta^I=- \bar{\eth}^2\eth^2  \mu^I
\end{equation}
\begin{equation} \label{e3}
2r\partial_r(r\partial_r \xi) 
+ \bar \eth \eth \xi =\bar{\eth}^2\eth^2 \mu^R.
\end{equation}

Since the right hand side of (\ref{e1}) has an expansion in spherical
harmonics with $l\geq 1$ and the right hand sides of (\ref{e2}),
(\ref{e3}) have expansions with $l\geq 2$, we can determine the expansion
coefficients of the unknowns for $l=0,1$. They can be given in the form
\[
\xi=A+rQ +\frac{1}{r} P,\quad
\eta^I=iJ + const.,\quad
\eta^R=rQ-\frac{1}{r} P + const.,
\]
with 
\begin{equation}
\label{eq:scon}
P=\frac{3}{2} P^an_a,  
\quad Q=\frac{3}{2}Q^an_a, 
\quad  J=3J^an_a,
\end{equation}
where $A, C^a, Q^a, J^a$ are arbitrary constants. 
Using (\ref{Psit}), we obtain the corresponding tensors in the form
(cf. (\cite{York})
\begin{align}
\label{eq:PsiPp}
\Psi_P^{ab}&=\frac{3}{2r^4}  \left( -P^a n^b-P^bn^a 
-(\delta^{ab}-5n^an^b)\,P^cn_c \right),\\
\label{eq:PsiJp}
\Psi_J^{ab}&=\frac{3}{r^3}(n^a  \epsilon^{bcd} J_c n_d +  
n^b\epsilon^{acd} J_c n_d),  \\
\label{eq:PsiAp}
\Psi_A^{ab}&=\frac{A}{r^3}\,(3n^an^b - \delta^{ab}),\\
\label{eq:PsiQp}
\Psi_Q^{ab}&=\frac{3}{2r^2}  \left( Q^a n^b+Q^bn^a 
- (\delta^{ab}-n^an^b)\,Q^cn_c \right). 
\end{align}

We assume now that $\xi$ and $\eta^I$ have expansions in terms of in
spherical harmonics with $l\geq 2$. Then there exists a smooth function
$\lambda_2$ of spin weight 2 such that
\[
\xi=\bar\eth^2 \lambda_2^R, \quad 
\eta_1^I=\bar \eth \lambda_2^I.
\] 
Using these expressions in equations (\ref{e1}) -- (\ref{e3})
and observing that for smooth spin weighted functions $\mu_s$ with $s>0$
we can have $\bar \eth \mu_s=0$ only if $\mu_s=0$, we obtain
\[
\eth \eta^R=-2r\partial_r\bar\eth \lambda_2^R, \quad
\eth^2 \mu^I=-r\partial_r \lambda_2^I,
\]
\[
\eth^2 \mu^R=2r\partial_r(r\partial_r\lambda_2^R)
-2\lambda_2^R+\eth \bar\eth\lambda_2^R.
\]
We are thus in a position to describe the general form of the
coefficients in the expression (\ref{Psit}) 
\begin{align}
\label{eq:xi}
\xi &= \bar\eth^2 \lambda_2^R + A + r\,Q + \frac{1}{r}\,P,\\
\label{eq:eta1}
\eta_1 &= -2\,r\,\partial_r\,\bar\eth \lambda_2^R
+ \bar\eth \lambda_2^I
+ r\,\eth Q - \frac{1}{r}\,\eth P 
+ i\,\eth J,\\
\label{eq:mu2}
\mu_2 &= 2\,r\,\partial_r(r\,\partial_r\,\lambda_2^R)-2\,\lambda_2^R
+ \eth \bar\eth\lambda_2^R-r\,\partial_r\,\lambda_2^I.
\end{align}

\begin{theorem} 
\label{t2}
Let $\lambda$ be an arbitrary smooth, complex, function in 
$B_a \setminus \{i\} \subset \mathbb{E}^3$ with 
$0 < a \le \infty$, and set $\lambda_2=\eth^2\lambda$. 
Then the tensor
\begin{equation}
\label{gensol}
\Psi^{ab} = \Psi_P^{ab} + \Psi_J^{ab}
+ \Psi_A^{ab} + \Psi_Q^{ab} + \Psi_{\lambda}^{ab},
\end{equation}
satisfies $D^a\Psi_{ab}=0$ in $B_a \setminus \{i\}$.
Here the first four terms on the right hand side are given by
(\ref{eq:PsiPp}) -- (\ref{eq:PsiAp}) while $\Psi_{\lambda}^{ab}$ 
is is obtained by using in (\ref{Psit}) only the part of the
coefficients (\ref{eq:xi}) -- (\ref{eq:mu2}) which depends on
$\lambda_2$. Conversely, any smooth solution in $B_a \setminus \{i\}$ of
the equation above is of the form (\ref{gensol}).
\end{theorem} 

Obviously, the smoothness requirement on $\lambda$ can be relaxed since
$\Psi_{\lambda}^{ab} \in C^1(B_a \setminus \{i\})$ if
$\lambda \in C^5(B_a \setminus \{i\})$. Notice, that no fall-off behaviour
has been imposed on $\lambda$ at $i$ and that it can show all kinds of
bad behaviour as $r \rightarrow 0$. 

Since we are free to choose $a$, we also obtain an expression for the
general smooth solution on $\mathbb{E}^3 \setminus \{i\}$. By suitable
choices of $\lambda$ we can construct solutions $\Psi_{\lambda}^{ab}$
which are smooth on $\mathbb{E}^3$ or which are smooth with compact
support.  

Given a subset S of $\mathbb{R}^3$ which is compact with boundary, we can
use the representation (\ref{gensol}) to construct hyperboloidal initial
data (\cite{Friedrich83}) on $S$ with a metric $h$ which is Euclidean on
all of $S$ or on a subset $U$ of S. In the latter case we would require
$\Psi_{\lambda}^{ab}$ to vanish on $S \setminus U$. In the case where the
trace-free part of the second fundamental form implied by $h$ on
$\partial S$ vanishes and the support of $\Psi^{ab}$ has empty
intersection with $\partial S$ the smoothness of the corresponding
hyperboloidal initial data near the boundary follows from the discussion
in (\cite{Andersson92}). Appropriate requirements on
$h$ and $\Psi^{ab}$ near $\partial S$ which ensure the smoothness of the
hyperboloidal data under more general assumptions can be found in
(\cite{Andersson94}).

There exists a 10-dimensional space of conformal Killing vector
fields on $\mathbb{E}^3$. In the cartesian coordinates $x^i$ a generic
conformal Killing vector field $\xi_0^a$ has components 
\begin{equation}
\label{eq:ckvflat2}
\xi_0^i= k^j\,(2\,x_j\,x^i - \delta_j\,^i\,x_l\,x^l)
+ \epsilon^i\,_{jk}\,S^j\,x^k + ax^i + q^i,
\end{equation}
where $k^i$, $S^i$, $q^i$ are arbitrary constant vectors and $a$ an
arbitrary number. In terms of the ``physical coordinates'' 
$y^i = x^i/|x|^2$,  with respect to which $i$ represents infinity, we see
that $k^i$, $S^i$, $q^i$, and $a$ generate translations, rotations,
``special conformal transformations'', and dilatations respectively.

For $0 < \epsilon < a$ we set $S_{\epsilon} = \{|x| = \epsilon\}$ and
denote by $dS_{\epsilon}$ the surface element on it. For the tensor field
$\Psi^{ab}$ of (\ref{gensol}) we obtain  
\begin{equation}
\label{eq:flatint}
\frac{1}{8\,\pi} \int_{S_{\epsilon}} 
\Psi^{ab}\,n_a\,\xi _{0b}\,dS_{\epsilon} =
(P^ak_a + J^aS_a + A\,a + Q^aq_a). 
\end{equation}
We note that the integral is independent of $\epsilon$ and, more
importantly, independent of the choice of $\lambda$. Thus the function
$\lambda$ does neither contribute to the momentum 
\begin{equation} 
\label{momentum}
P^a= \frac{1}{8\pi} \lim_{\epsilon \to 0 } \int_{S_{\epsilon}} 
r^2\,\Psi_{bc}\,n^b\, (2n^cn^a - \delta^{ca})\,dS_{\epsilon},
\end{equation}
nor to the angular momentum
\begin{equation} 
\label{angmomentum}
J^a= \frac{1}{8\pi} \lim_{\epsilon\to 0 } 
\int_{S_{\epsilon}} r\,
\Psi_{bc}\,n^b \epsilon^{cad}\,n_d\,dS_{\epsilon},
\end{equation}
of the data.

If we use the coordinates $x^i$ to identify $\mathbb{E}^3$ with
$\mathbb{R}^3$ and map the unit sphere $S^3 \subset \mathbb{R}^4$ by
stereographic projection through the south pole onto $\mathbb{R}^3$, the
point 
$i$, i.e. $x^i = 0$, will correspond to the north pole, which will
be denoted in the following again by $i$. The south pole, denoted in the
following by $i'$, will represent infinity with respect to the coordinates
$x^i$ and the origin with respect to the coordinates $y^i$. We use $x^i$
and $y^i$ as coordinates on $S^3 \setminus \{i'\}$ and 
$S^3 \setminus \{i\}$ respectively. If $h^0$ is the standard metric on
$S^3$, we have in the coordinates $x^i$
\begin{equation}
\label{eq:csf}
h^0_{kl}=\theta^{-2} \delta_{kl}, \quad
\theta= \left(\frac{1+r^2}{2} \right)^{1/2}.
\end{equation}

We assume that the function $\lambda$ is smooth in
$\mathbb{E}^3 \setminus \{i\}$ and set 
$\tilde \Psi^{ab} = \theta^{10}\,\Psi^{ab}$ with $\Psi_{ab}$ as in
(\ref{gensol}). Then, by (\ref{eq:resc}), (\ref{eq:cresc}),
\begin{equation}
\label{eq:psis3}
D_a \tilde{\Psi}^{ab}=0 \text{ in } S^3 \setminus \{i, i'\},
\end{equation}
where $D_a$ denotes the connexion corresponding to $h^0$. 
The smoothness of $\tilde{\Psi^{ab}}$ near $i$ can be read off from 
equations (\ref{Psit}) and (\ref{eq:xi}) -- (\ref{eq:mu2}). In order to
study its smoothness near $i'$ we perform the inversion to obtain the
tensor in the coordinates $y^i$. It turns out that we obtain the same
expressions as before if we make the replacements
\[
n^i \rightarrow -n^i, \quad  
m^i \rightarrow m^i,
\]
\[
r \rightarrow 1/r, \quad  
\xi \rightarrow \xi, \quad 
\eta_1 \rightarrow -\eta_1, \quad 
\mu \rightarrow \mu,
\]
\[
P^a\rightarrow Q^a, \quad 
J^a \rightarrow -J^a, \quad  
A \rightarrow A, \quad
Q^a \rightarrow P^a.
\]
Thus the tensors (\ref{eq:PsiPp}) -- (\ref{eq:PsiAp}), the first two of
which are the only ones which contribute to the momentum and angular
momentum, are singular in $i'$ as well as in $i$. Observing again the
conformal covariance (\ref{eq:resccko}), we obtain the following result.  

\begin{corollary}
\label{psis3}
The general smooth solution of the equation $D_a\tilde{\Psi}^{ab}=0$ on 
$S^3 \setminus \{i, i'\}$ with respect to the metric $\omega^4\,h^0$,
where $h^0$ denotes the standard metric on the unit $3$-sphere and 
$\omega \in C^{\infty}(S^3)$, $\omega > 0$, is given by
$\tilde{\Psi}^{ab} = (\omega^{-1}\,\theta)^{10}\,\Psi_{ab}$ with
$\Psi_{ab}$ as in (\ref{gensol}) and $\lambda \in C^{\infty}(\mathbb{E}^3
\setminus
\{i\})$.  If we require the solution to be bounded near $i'$ (in
particular, if we construct solutions with only one asymptotically
flat end), the quantities $P^a$, $J^a$, $A$, $Q^a$ must vanish.
\end{corollary}

We can now provide tensor fields which satisfy
condition (\ref{condd}) and thus prove a special case of theorem
\ref{T2}.

\begin{theorem} 
\label{flatang}
Denote by $\Psi^{ab}$ a tensor field of the type (\ref{gensol}).
If $r\lambda \in  E^\infty(B_a)$ and $P^a=0$, then  
$r^8\Psi_{ab}\Psi^{ab} \in E^\infty(B_a)$.
\end{theorem}

Only the part of $\lambda$ which in an expansion in terms of spherical
harmonics is of order $l\geq 2$ contributes to $\Psi_{ab}$. We note 
that the condition $r\lambda \in  E^\infty(B_a)$ entails that
this part is of order $r$. The singular parts of $\Psi_{ab}$ 
are of the form (\ref{eq:PsiJp}) -- (\ref{eq:PsiAp}).

For the proof we need certain properties of the $\eth$ operator. If $m^a$
is suitably adapted to standard spherical coordinates, the $\eth$
operator, acting on functions of spin weight $s$, acquires the form
\begin{equation}
\label{eq:eth}
\eth \eta_s = -(\sin\theta)^s \left [\frac{\partial}{\partial
\theta}+\frac{i}{\sin\theta}  \frac{\partial}{\partial
\phi} \right]\left((\sin\theta)^{-s} \eta_s \right).
\end{equation}
The operator $\bar \eth\eth$ acting on functions with $s=0$ is the
Laplace operator on the unit sphere, i.e. we have the identity
\begin{equation}
\label{eq:ls2}
\bar \eth \eth f=r^2\Delta f - x^i x^j \partial_i \partial_j f -2
x^i\partial_i f, 
\end{equation}
where $\Delta$ denotes the Laplacian on $\mathbb{R}^3$.
The commutator is given by 
\begin{equation}
\label{eq:com}
  (\bar \eth \eth-\eth \bar \eth)\eta_s =2s\eta_s.
\end{equation}
From this formula we obtain for $q \in \mathbb{N}$ 
by induction the relations
\begin{equation}
\label{eq:comp}
  (\bar \eth \eth^q -\eth^q \bar \eth)\eta_s
  =(2qs+(q-1)q)\eth^{q-1}\eta_s, 
\end{equation}
\begin{equation}
\label{eq:compb}
  ( \eth \bar\eth^q -\bar \eth^q  \eth)\eta_s
  =(-2qs+(q-1)q)\eth^{q-1}\eta_s. 
\end{equation}
In particular, it follows that the functions $\eta$ and $\mu$, which
satisfy $\eth \eta =\eta_1$ and $\eth^2\mu =\mu_2$, are given by
\begin{align}
  \label{eq:eta}
  \eta &=-2r \partial_r (\bar \eth \eth \lambda^R +2\lambda^R )+\bar
  \eth \eth \lambda^I +2\lambda^I+rQ-\frac{P}{r}+iJ,\\
\mu &=2r \partial_r (r \partial_r \lambda^R) +\bar\eth \eth \lambda^R
  -2r \partial_r \lambda^I \label{eq:mu}.
\end{align}

\begin{lemma} 
\label{eth}
Let $f, g\in E^\infty$ be two complex functions (spin weight zero).
Then, the functions $\bar\eth^q f\eth^q g$, $q \in \mathbb{N}$,  
of spin weight zero are also in $E^\infty$.
\end{lemma}

Remarkably, the statement is wrong if we replace 
$E^\infty$ by $C^\infty$: the calculation in polar coordinates, using 
(\ref{eq:eth}), gives 
\[
\eth x^1\, \bar \eth x^3 = - x^1\,x^3 + i\,r\,x^2.
\]

\noindent
{\bf Proof:}
For $q=1$ the proof follows from two identities. The first one is a simple
consequence of the Leibnitz rule
\begin{equation}
\label{eq:leib}
\eth f \bar \eth g+ \bar \eth f  \eth g=\bar \eth \eth(fg)
-f \bar \eth \eth g -g \bar \eth \eth f.
\end{equation}
Since $\bar \eth \eth $ maps by (\ref{eq:ls2}) smooth functions into
smooth functions, it follows that
$(\eth f \bar \eth g+ \bar \eth f  \eth g) \in E^\infty$ if
$f,g \in E^\infty$ (here we can replace $E^\infty$ by $C^\infty$).

The other identity reads
\begin{equation}
\label{eq:i}
\eth f \bar \eth g- \bar \eth f  \eth g=2\,i\,r\,\epsilon_l\,^{jk}\,x^l
\,\partial_j f \,\partial_k g.
\end{equation}
It is obtained by expressing (\ref{eq:eth}) in the Cartesian coordinates
$x^l$.  Important for us is the appearance of the factor $r$.  A
particular case of this relation has been derived in \cite{Held70}. It
follows that 
$(\eth f \bar \eth g- \bar \eth f  \eth g) \in E^\infty$ if 
$f,g \in E^\infty$. Taking the difference of (\ref{eq:leib}) and
(\ref{eq:i}) gives the desired result.

To obtain the result for arbitrary $q$, we proceed by induction. 
The Leibniz rule gives
\[
\eth^{q+1}f \bar \eth^{q+1} g =\bar \eth \eth (\eth^q f \bar \eth^q g
)-\bar \eth \eth^q f \eth \bar \eth^q g - \eth^q f \eth \bar
\eth^{q+1} g - \bar \eth ^q g \eth \bar \eth \eth^q f. 
\]
The induction hypothesis for $q$ and  (\ref{eq:ls2}) imply that 
the first term on the right hand side is in $E^{\infty}$.
The factors appearing in the following terms can be written by
(\ref{eq:comp}) and  (\ref{eq:compb}) in the form
\[
\bar \eth \eth^q f = \eth^{q-1}(\eth \bar \eth f +q(q-1)f  ),\quad
\eth \bar\eth^q g = \bar \eth^{q-1}(\eth \bar \eth g +q(q-1)g  ),
\]
\[
\eth \bar\eth^{q+1} g = \bar \eth^q(\bar \eth \eth g +q(q-1) g  ),\quad
\eth \bar \eth \eth^q f = \eth^q(\eth \bar \eth f +q(q-1)f  ).
\]
Since the functions in parenthesis are, by (\ref{eq:ls2}), in $E^\infty$,
the induction hypothesis implies that each of the products  
is in $E^\infty$. 
$\blacksquare$\\

\noindent
{\bf Proof of theorem \ref{flatang}:}
In terms of the coefficients (\ref{eq:xi}) -- (\ref{eq:mu2}) we have
\begin{equation} 
\label{Psi^2}
r^8\Psi_{ab}\Psi^{ab}=r^2 \left(2 \mu_2 \bar\mu_2 +2\eta_1\bar\eta_1 
+3 \xi^2 \right). 
\end{equation}
Since   $r\lambda$ is in $E^\infty(B_a)$, equations 
(\ref{eq:eta}) and (\ref{eq:mu}) imply that 
$r\xi, r\eta, r\mu \in E^\infty$. 
The conclusion now follows from lemma \ref{eth}. 
$\blacksquare$

\subsection{The general case: existence} 
\label{existencemom}

The existence of solution to the momentum constraint for asymptotically 
flat initial data has been proved in weighted Sobolev spaces 
(cf.  \cite{Cantor}, \cite{Choquet99}, \cite{Choquet80},
\cite{Christodoulou81} and the reference given there)
and in weighted H\"older spaces \cite{Chaljub-Simon82}. The existence of
initial data with non trivial momentum and angular momentum and the role
of conformal symmetries have been analysed in some detail in
\cite{Beig96}. In this section we will prove existence of solutions to the
momentum constraint with non-trivial momentum  and angular momentum
following the approach of \cite{Beig96}. We generalize some of the
results shown in \cite{Beig96} to  metrics in the class (\ref{eq:sobh}). 
The results of the previous section will be important for the analysis of
the general case. 

We use the York splitting methods to reduce the problem of solving the
momentum constraint to solving a linear elliptic system of equations.
Let the conformal metric $h$ on the initial hypersurface $S$ be given. We
use it to define the overdetermined elliptic conformal Killing operator
$\mathcal{L}_h$, which maps vector fields $v^a$ onto symmetric
$h$-trace-free tensor fields according to    
\begin{equation}
 \label{eq:calL}
(\mathcal{L}_h v)^{ab} = D^av^b+D^bv^a - \frac{2}{3}\,h^{ab}\,D_c v^c,
\end{equation}
and the underdetermined elliptic divergence operator $\delta_h$ which maps
symmetric
$h$-trace-free tensor fields $\Phi^{ab}$ onto vector fields according to 
\begin{equation}
\label{eq:delh}
(\delta_h \Phi)^a = D_b\Phi^{ba}.
\end{equation}
Let $\Phi^{ab}$ be a symmetric $h$-trace-free tensor field and set
\begin{equation}
\label{eq:split}
\Psi^{ab}= \Phi^{ab}-(\mathcal{L}_h v)^{ab}.
\end{equation}
Then $\Psi^{ab}$ will satisfy the equation $D_a\Psi^{ab}=0$ if the vector
field $v^a$ satisfies 
\begin{equation}
  \label{eq:ev}
\mathbf{L}_hv^a=D_b \Phi^{ab},  
\end{equation}
where the operator $\mathbf{L}_h $ is given by
\begin{equation} 
\label{eq:Lhs}
\mathbf{L}_h v^a =D_b (\mathcal{L}_h v)^{ab}
= D_b D^b v^a +\frac{1}{3}D^a D_b v^b +R^a\,_b v^b.  
\end{equation}
Since  
$\Psi^{ab}(\mathcal{L}_h v)_{ab} = 2\,D_a(\Psi^{ab} v_b)
- 2\,(\delta_h\Psi)^{a} v_a$
for arbitrary vector fields $v^a$ and symmetric $h$-trace-free
tensor fields $\Psi^{ab}$, $\mathcal{L}_h$ has formal adjoint
$\mathcal{L}_h^* = - 2\,\delta_h$. Thus   
\begin{equation}
\label{eq:lh*lh}
\mathbf{L}_h = - \frac{1}{2}\,\mathcal{L}_h^* \circ \mathcal{L}_h,
\end{equation}
and the operator is seen to be elliptic. 

Provided the given data are sufficiently smooth, we can use theorem
\ref{Cantor} to show the existence of solutions to (\ref{eq:ev}). 

\begin{lemma}[Regular case]
\label{regularcase}
Assume that the metric satisfies (\ref{eq:sobh}) and  
$\Phi^{ab}$ is an $h$-trace-free  symmetric tensor field in 
$W^{1,p}(S)$, $p>1$. Then there exists a unique vector field 
$v^a\in W^{2,p}(S)$ such that the tensor field
$\Psi^{ab}= \Phi^{ab}-(\mathcal{L}_h v)^{ab}$ satisfies the equation
$D_a\Psi^{ab}=0$ in $S$.
\end{lemma}

\noindent
{\bf Proof:} The well known argument that the condition of theorem
\ref{Cantor} will be satisfied extends to our case. Assume that the
vector field $\xi^a$ is in the kernel of $L_h$, i.e. 
\begin{equation}
\label{eq:ckv}
\mathbf{L}_h \xi^a=0 \text{ in } S.
\end{equation}
Since the metric satisfies (\ref{eq:sobh}), elliptic regularity gives
\begin{equation}
\label{eq:rckv}
\xi^a\in C^{2,\alpha}(S).
\end{equation}
This smoothness suffices to conclude from
(\ref{eq:lh*lh}), (\ref{eq:ckv}) that
$0 = - 2\,(\xi, \mathbf{L}_h \xi)_{L^2} 
= (\mathcal{L}_h \xi,\mathcal{L}_h \xi)_{L^2}$, whence   
\begin{equation}
\label{eq:ckv2}
(\mathcal{L}_h \xi)_{ab}=0.  
\end{equation}
This implies for an arbitrary symmetric, $h$-trace-free tensor field 
$\Phi^{ab} \in W^{1,p}(S)$, $ p > 1$, the relation
$0 = (\mathcal{L}_h \xi, \Phi)_{L^2} 
= - 2\,(\xi, \delta_h \Phi)_{L^2}$,
which shows that the Fredholm condition will be satisfied for
any choice of $\Phi^{ab}$ in (\ref{eq:ev}).
$\blacksquare$\\

We call the case above the ``regular case'' because the solution still
satisfies the condition $\Psi^{ab} \in W^{1,p}(S)$. While this allows us
to have solutions diverging like $O(r^{-1})$ at given points, it excludes
solutions with non-vanishing momentum or angular momentum.

We note that by (\ref{eq:ckv2}) the kernel of $\mathbf{L}_h$ consists of
conformal Killing fields. Let $\xi^a$ be such a vector field. Using
(\ref{eq:rckv}) and lemma \ref{holderrest} we find that we can write in
normal coordinates centered at the point $i$ of $S$ 
\begin{equation}
\label{eq:eckv}
\xi^k= \xi_0^k + O(r^{2+\alpha}),
\end{equation}
where $\xi_0^k$ is the ``flat'' conformal Killing field 
(\ref{eq:ckvflat2}) with coefficients given by
\begin{equation}
\label{eq:ckvdata}
k_a = \frac{1}{6}\,D_a D_b \xi^b(i), \quad
S^a={\epsilon^a}_{bc}D^b \xi^c(i), \quad  
q^a=\xi^a(i), \quad  
a =\frac{1}{3} D_a\xi^a(i). 
\end{equation}
Since $S$ is connected, the integrability conditions for conformal Killing
fields (cf. \cite{Yano57}) entail that these ten ``conformal Killing data
at $i$'' determine the field $\xi^a$ uniquely on $S$. 

With a conformal rescaling of the metric with a smooth, positive,
conformal factor $\omega$ 
\begin{equation}
\label{eq:resc}
h_{ab} \rightarrow h'_{ab}=\omega^4 h_{ab},
\end{equation}
which implies a transition of the connection $D_a \rightarrow D'_a$,
we associate the rescalings
\begin{equation}
\label{eq:cresc}
\Psi^{ab} \rightarrow {\Psi'}^{ab}=\omega^{- 10} \Psi^{ab}, \quad 
\xi^a \rightarrow {\xi'}^a =  \xi^a,
\end{equation}
for $h$-trace free, symmetric tensor fields $\Psi^{ab}$ and Killing fields
$\xi^a$. Then the conformal Killing operator and the divergence operator
satisfy
\begin{equation}
\label{eq:resccko}
(\mathcal{L}_{h'}v)^{ab}=\omega^{-4}(\mathcal{L}_{h}v)^{ab}, \quad
D'_a {\Psi'}^{ab} = \omega^{- 10} D_a \Psi^{ab}.
\end{equation}
If we write $\omega=e^{f}$, the conformal Killing data transform as
\begin{align}
{k'}^a &= k^a+2a D^af(i)+\epsilon^{abc} S_b D_c f(i)+ q_c D^aD^c f(i),
\label{eq:k'}\\ 
{S'}^a &= S^a+2\epsilon^{abc}q_b D_c f(i), \label{eq:S'}\\
a'&=a+q^aD_af(i) \label{eq:a'} \\
{q'}^a &= q^a  \label{eq:q'}.
\end{align}

A vector field $\xi$ on $S$ will be called an ``asymptotically conformal
Killing field'' at the point $i$ if it satisfies in normal coordinates
$x^k$ centered at $i$ equation (\ref{eq:eckv}) with (\ref{eq:ckvflat2}).
While any conformal Killing field is an asymptotically conformal Killing
field, the converse need not be true.  Let $\xi^a$ be an asymptotically 
conformal Killing field and consider the integral
\begin{equation}
\label{eq:coint}
I_{\xi}=\frac{1}{8\pi}\lim_{\epsilon \rightarrow 0}
\int_{\partial B_\epsilon} \Psi^{ab}\,n_a\,\xi_b\,dS_\epsilon,
\end{equation}
where in the coordinates $x^k$ the unit normal $n^k$ to the sphere
$\partial B_\epsilon$ around $i$ approaches $x^k/|x|$ for small
$\epsilon$. As shown in the previous section, the integral vanishes if the
tensor $\Psi^{ab}$ is of order $r^{-1}$ at $i$. If, however,
$\Psi^{ab}$ is of order $r^{-n}$, $n=2,3,4$, at $i$, the integral gives
non-trivial results and can be understood in particular as a linear form
on the momentum and angular momentum of the data.

We recall two important properties of the integral $I_\xi$. The first is
the fact that the integral $I_\xi$ is invariant under the rescalings
(\ref{eq:resc}), (\ref{eq:cresc}). The second property is concerned with
the presence of conformal symmetries. Let $\Phi^{ab}$ be an arbitrary
$h$-trace-free tensor field and $v^a$ an arbitrary vector field.  Using 
Gauss' theorem, we obtain
\begin{equation}
\label{eq:Gauss}
2\,\int_{\partial B_\epsilon} \Phi^{ab}\,n_a\,v_b\, dS_\epsilon
= - 2\,\int_{S-B_\epsilon} v_a\,D_b \Phi^{ab} \, d\mu
- \int_{S-B_\epsilon}(\mathcal{L}_hv)_{ab} \Phi^{ab}\,d\mu,
\end{equation}
with the orientation of $n^k$ as above. If we apply this equation to
(\ref{eq:coint}), the first term on the right hand side will vanish if
$D_a\Psi^{ab}=0$ in $S\setminus \{i\}$,  while the second term on the
right hand side need not vanish if $\xi^a$ is only an
asymptotically conformal Killing vector field. However, if $\xi$ is a
conformal Killing field, we get $I_\xi = 0$. Thus the presence of Killing
fields entails restrictions on the values allowed for the momentum
and angular momentum of the data. For this reason the presence of
conformal symmetries complicates the existence proof. Note that we are
dealing with vector fields $\xi^a$ which satisfy the conformal Killing
equation (\ref{eq:ckv2}) everywhere in $S$; in general, a small local
perturbation of the metric will destroy this conformal symmetry.

Observing the conformal covariance (\ref{eq:resccko}) of
the divergence equation, we perform a rescaling of the form
(\ref{eq:cf}), (\ref{eq:h00}) such that the metric $h'$ has in $h'$-normal
coordinates $x^k$ in $B_a$ centered at $i$ local expression $h'_{kl}$ with
\begin{equation}
\label{eq:hn3exp}
h'_{kl} - \delta_{kl} = O(r^3), \quad
\partial_j h'_{kl} = O(r^2).
\end{equation}
In these coordinates let 
$\Psi^{ik}_{flat}\in C^\infty(B_a\setminus \{i\})$ 
be a trace free symmetric and divergence free tensor with respect
to the flat  metric $\delta_{kl}$, 
\begin{equation}
\label{eq:psi0}
\delta_{ik}\,\Psi^{ik}_{flat} = 0, \quad  
\partial_i\Psi_{flat}^{ik}=0 \text{ in } B_a\setminus \{ i\},
\end{equation}
with
\begin{equation}
\label{eq:o-4}
\Psi_{flat}^{ik} = O(r^{- 4}), \quad
\partial_j\Psi_{flat}^{ik} = O(r^{- 5})
\text{ as } r\rightarrow 0. 
\end{equation}
All these tensors have been described in theorem (\ref{t2}).
Note that the conditions (\ref{eq:o-4}) are essentially conditions on the
function $\lambda$ which characterizes the part
$\Psi_{\lambda}^{ab}$ of (\ref{gensol}). Denote by 
$\Phi^{ab}_{sing} \in C^\infty(S \setminus \{ i\})$ 
the $h'$-trace free tensor which is given on 
$B_a \setminus \{i\}$ by
\begin{equation}
\label{eq:Qn}
\Phi^{ab}_{sing}= \chi\,(\Psi_{flat}^{ab}-\frac{1}{3}\,
h'_{cd}\,\Psi_{flat}^{cd}\,{h'}^{ab}),
\end{equation}
and vanishes elsewhere. Here $\chi$ denotes a smooth function of
compact support in $B_a$ equal to $1$ on $B_{a/2}$. By our assumptions we
have then
\begin{equation}
\label{eq:dviPhi-2}
D'_a\Phi^{ab}_{sing} = O(r^{-2}) \text{ as } \rightarrow 0.
\end{equation}

\begin{theorem}[Singular case]
\label{sing}
Assume that the metric $h$ satisfies (\ref{eq:sobh}), 
$\omega_0$ denotes the conformal factor (\ref{eq:cf}),
$h' = \omega_0^4\,h$,
$\Phi^{ab}_{sing}$ is the tensor field defined above, and
$\Phi_{reg}^{ab}$ is a symmetric $h$-trace free tensor field
in $W^{1,p}(S)$, $p>1$.\\ 

\noindent
i) If the metric $h$ admits no conformal Killing fields on $S$, then
there exists a unique vector field $v^a \in W^{2, q}(S)$, with $q=p$
if $p < 3/2$ and $1 < q < 3/2$ if $p \geq 3/2$, such that the tensor field
\begin{equation}
\label{eq:singe}
\Psi^{ab} = \omega_0^{10}\,(\Phi_{sing}^{ab} + \Phi_{reg}^{ab} 
+ (\mathcal{L}_{h'} v)^{ab}),
\end{equation}
satisfies the equation $D_a \Psi^{ab}=0$ in $S\setminus \{i\}$.\\

\noindent
ii) If the metric $h$ admits conformal Killing fields $\xi^a$ on $S$,
a vector field $v^a$ as specified above exists if and only if the
constants $k^a$, $S^a$, $a$, $q^a$ (partly) characterizing the
tensor field $\Phi_{sing}^{ab}$ (cf. (\ref{gensol}), 
(\ref{eq:PsiPp})--(\ref{eq:PsiQp})), satisfy the equation
\begin{equation}
\label{eq:sycond}
P^a\,k_a + J^a\,S_a + A\,a + (P^c\,L_c\,^a(i) + Q^a)\,q_a = 0,    
\end{equation}
for any conformal Killing field $\xi^a$ of $h$, where the constants 
$k_a$, $S_a$, $a$, $q_a$ characterizing $\xi^a$ are given by 
(\ref{eq:ckvdata}).\\    

In both cases the momentum and angular momentum (cf. (\ref{momentum}),
(\ref{angmomentum})) of $\Psi^{ab}$ agree with those of the tensor
$\Phi_{sing}^{ab}$. These quantities can thus be prescribed freely in 
case (i).
\end{theorem}

\noindent
{\bf Proof:}
Because of (\ref{eq:dviPhi-2}) we can consider 
$D'_a (\Phi_{sing}^{ab} + \Phi_{reg}^{ab})$ as a function in 
$L^p(S)$, $1 < p < 3/2$. In case (i) the kernel of the operator
$\mathcal{L}_{h'}$ appearing in the equation
\[
D'_{a}(\Phi_{sing}^{ab} + \Phi_{reg}^{ab} 
+ (\mathcal{L}_{h'} v)^{ab}) = 0,
\] 
is trivial and we can apply theorem \ref{Cantor} to show that the
equation above determines a unique vector field $v^a$ with the
properties specified above.  After the rescaling we have $D_a\Psi^{ab}=0$
by (\ref{eq:resccko}).

In case (ii) the kernel of $\mathcal{L}_{h'}$ is generated by the
conformal Killing fields ${\xi'}^a  = \xi^a$ of $h'$. If we express
(\ref{eq:Gauss}) in terms of the metric $h'$, the tensor field
$\Phi_{sing}^{ab} + \Phi_{reg}^{ab}$, and the vector field 
${\xi'}^a$, take the limit $\epsilon \rightarrow 0$ and use equation
(\ref{eq:flatint}), we find
that the Fredholm condition of theorem \ref{Cantor} is satisfied if and
only if for every conformal Killing field ${\xi'}^a$ of $h'$ we have
\[
P^a\,k'_a + J^a\,S'_a + A\,a' + Q^a\,q'_a = 0,    
\]
where the constants $k'_a$, $S'_a$, $a'$, $q'_a$ are given by equation    
(\ref{eq:ckvdata}), expressed in terms of ${\xi'}^a$ and $h'$. By 
(\ref{eq:k'}) -- (\ref{eq:q'}) this condition is identical with 
(\ref{eq:sycond}).

As shown in the previous section, we can choose $\Phi_{sing}^{ab}$ such
that the corresponding momentum and angular momentum integrals take
preassigned values, which can be chosen freely in case (i) and need to
satisfy (\ref{eq:sycond}) in case (ii). These values will agree with
those obtained for $\omega_0^{-10} \Psi^{ab}$ due to the regularity
properties of $v^a$. After the rescaling the values of the momentum and
the angular momentum remain unchanged because $\omega_0 = 1 + O(r^2)$.
$\blacksquare$\\

We note that it is the presence of the $10$-dimensional space of conformal
Killing fields on the standard $3$-sphere which led to the observation
made in corollary (\ref{psis3}). The latter generalizes as follows. 

\begin{corollary}
\label{corsy}
If $S$ is an arbitrary compact manifold, $h$ satisfies (\ref{eq:sobh}),
and we allow for $p \ge 2$ asymptotic ends $i_k$, $1 \le k \le p$, 
we can choose the sets of constants $(P^a_k, J^a_k, A_k, Q^a_k)$
arbitrarily in the ends $i_k$, $1 \le k \le p - 1$. Which constants can
be chosen at the end $i_p$ depends on the  conformal Killing fields
admitted by $h$.
\end{corollary}

\noindent
{\bf Proof:}
This follows from the observation that in the case of $p$ ends equation
(\ref{eq:sycond}) generalizes to  an equation of the form 
\[
\sum_{l = 1}^p
P^a_l\,k_a^l + J^a_l\,S_a^l + A_l\,a^l 
+ (P^c_l\,L_c\,^a(i_l) + Q^a_l)\,q^l_a = 0,    
\]
where the constants bear for given $l$ the same meaning with respect
to the point $i_l$ as the constants in (\ref{eq:sycond}) with
respect to $i$.  
$\blacksquare$\\

The case of spaces conformal to the unit $3$-sphere $(S^3,h_0)$ is very
exceptional. A result of Obata \cite{Obata71}, discussed in \cite{Beig96}
in the context of the constraint equations, says that unless the
manifold $(S,h)$ is conformal to $(S^3,h_0)$ there exists a smooth
conformal factor such that in the rescaled metric $h'$ every conformal
Killing field is in fact a Killing field. Thus the dimension of the space
of conformal Killing fields cannot exceed $6$. In fact, it has been shown
in \cite{Beig96} that in that case $h'$ can admit at most four
independent Killing fields and only one of them can be a rotation.
In this situation equation (\ref{eq:sycond}), written in terms of the
metric $h'$, reduces by (\ref{eq:ckvdata}) to
\[
J^a\,S'_a  +  (Q^a + P^b\,L'_a\,^b(i))\,q_a =0,  
\]
since $D'_a \xi^a = 0$ for a Killing field. The constants $P^a$
and $A$ can be prescribed arbitrarily. If there does exist a rotation
among the Killing fields, the equation above implies
\[
J^a\,S'_a = 0, \quad  (Q^a + P^b\,L'_a\,^b(i))\,q_a = 0.  
\]

\subsection{Asymptotic expansions near $i$ of solutions to the momentum
constraint} 
\label{asymtoticmomentum}

In this section we shall prove an analogue of Theorem \ref{EL} 
for the operator $\mathbf{L}_h$ defined in (\ref{eq:Lhs}). It will be used
to analyse the behaviour of the solutions to the momentum constraint
considered in theorem (\ref{sing}) near $i$ and to show the existence of
a general class of solutions which satisfy condition (\ref{condd}). Our
result rests on the close relation between the operator
$\mathbf{L}_h$ and the Laplace operator. 

We begin with a discussion on $\mathbb{R}^3$ and write  
$x_i = x^i$, $\partial^i = \partial_i$. The flat space analogue of
$\mathbf{L}_h$ on $\mathbb{R}^3$ is given by
\begin{equation}
\label{eq:L0}
\mathbf{L}_0 v^k= \Delta v^k +\frac{1}{3}\,
\partial^k\,\partial_l\,v^l,
\end{equation}
where $\Delta$ denotes the flat space Laplacian and $v^k$ a vector field
on some neighbourhood of the origin in $\mathbb{R}^3$.

The following spaces of vector fields whose components are homogenous
polynomials of degree $m$ and smooth functions respectively will be
important for us.
\begin{definition}
\label{QmQinfty}
Let $m\in \mathbb{N}$, $m\geq1$. We define the real vector spaces 
$\mathcal{Q}_m$, $\mathcal{Q}_\infty (B_a)$ by 
\[
\mathcal{Q}_m=\{ v \in C^{\infty}(\mathbb{R}^3, \mathbb{R}^3)
|\, v^i \in \mathcal{P}_m, 
\,\,\, v^ix_i=r^2 v \text{ with } v\in \mathcal {P}_{m-1}  \},
\]
\[
\mathcal{Q}_\infty (B_a)=\{v \in C^{\infty}(B_a, \mathbb{R}^3)
|\,v^ix_i=r^2 v \text{ with } v \in C^\infty (B_a) \}.
\]
\end{definition}
The following lemma, an analogue of lemma \ref{lE_M},
rests on the conditions imposed on the vector fields above. 

\begin{lemma} 
\label{L0P}
Suppose $s\in \mathbb{Z}$. Then the operator $\mathbf{L}_0$ defines a
linear, bijective mapping of vector spaces
\[
\mathbf{L}_0: r^s \mathcal {Q}_m \rightarrow r^{s-2} \mathcal {Q}_m,
\]
in the following cases:

(i) $s>0$

(ii) $s<0$, $|s|$ is odd and  $m+s\geq 0$.
\end{lemma}

Note that the assumptions on $m$ and $s$ imply that the vector field 
$\mathbf{L}_0(r^s\,p^i) \in C^{\infty}
(\mathbb{R}^3 \setminus \{ 0 \}, \mathbb{R}^3)$ defines a vector field in
$L^1_{loc}(\mathbb{R}^3,\mathbb{R}^3)$ which represents 
$\mathbf{L}_0(r^s\,p^i)$ in the distributional sense.\\

\noindent
{\bf Proof:}
For $s$ as above and $p^i \in \mathcal {Q}_m$ there exists 
some $p_{m-1}\in \mathcal{P}_{m-1}$ with 
\begin{equation} 
\label{qm-1}
\partial_k(r^s\,p^k) = r^s\,q_{m-1} 
\text{ with }
q_{m-1} = s\,p_{m-1} + \partial_k\,p^k \in \mathcal{P}_{m-1}.  
\end{equation}
With equation (\ref{DP}) it follows that
\[
\mathbf{L}_0( r^s p^i)=r^{s-2} \hat p^i \text{ with }
\]
\[
\hat p^i = s\,(s+1+2m)\,p^i + r^2\,\Delta p^i 
+ \frac{1}{3}\,(s\,x^i\,q_{m-1} 
+ r^2\,\partial^i q_{m-1}) \in \mathcal{P}_{m}. 
\]
Moreover, $\hat p^i \in \mathcal{Q}_m$
because $\hat{p}^i\,x_i = r^2\,\hat{p}_{m - 1}$ with 
\[
\hat{p}_{m - 1} = s\,(s+1+2m)\,p_{m-1} + x_i\,\Delta p^i 
+ \frac{1}{3}(s\,q_{m-1} + x_i\,\partial^i q_{m-1}) \in \mathcal{P}_{m-1}.
\]

To show that the kernel of the map is trivial, assume that 
$\mathbf{L}_0(r^s p^i) = 0 \in L^1_{loc}(\mathbb{R}^3, \mathbb{R}^3)$.
Taking a (distributional) derivative we obtain
\begin{equation} 
\label{eqm}
0 = \partial_i\,\mathbf{L}_0(r^s p^i) = 
\frac{4}{3}\,\Delta(\partial_i(r^s\,p^i)) = 
\frac{4}{3}\,\Delta(r^s\,q_{m-1}).
\end{equation} 
When $s>0$ or $|s|$ odd and $m-1+s\geq 0$ we use lemma \ref{lE_M} to 
conclude that $q_{m-1}=0$. We insert this in the equation 
$\mathbf{L}_0(r^s p^i)=0$ to obtain $\Delta (r^s p^i)=0$ and conclude
again by lemma \ref{lE_M} that $p^i=0$. 

There remains the case $s+m=0$ with $|s|$ odd.
Expanding $q_{m-1}$ in equation (\ref{eqm}) in harmonic polynomials 
(cf. (\ref{HH})), we get
\[
0 = \Delta  (r^{-m} q_{m-1})=\sum_{0\leq k\leq (m-1)/2}  
\Delta(r^{2k-m}h_{m-1-2k}),
\]
whence, by (\ref{DP}),
\[
\sum_{0\leq k\leq (m-1)/2} r^{2k - m - 2}\,
(m - 2\,k)\,(m - 2\,k - 1)\,h_{m-1-2k}=0.
\]
Since this sum is direct each summand must vanish
separately. Since $m$ is odd, the only factor
$(m - 2\,k - 1)$ which 
vanishes occurs when $2k = m-1$, from which we conclude that  
$q_{m-1}=r^{m-1}h_0$ with a constant $h_0$. Since equation (\ref{eqm}),
which reads now $h_0\,\Delta(r^{-1}) = 0$, holds in the distributional
sense, it follows that $h_0 = 0$. 
$\blacksquare$\\ 

Unless noted otherwise we shall assume in the following that the metric
$h$ is of class $C^{\infty}$ and that it is chosen in its conformal class
such that its Ricci tensor vanishes at $i$ (cf. (\ref{eq:cf}),
(\ref{eq:h00})). By $x^i$ will always be denoted a system of
$h$-normal coordinates centered at $i$ and all our calculations will be
done in these coordinates. Thus we have  
\[
h_{kl} = \delta_{kl} + O(r^3), \quad
\partial_j h_{kl} = O(r^2).
\]
We write the operator $\textbf{L}_h$ in the form
\[
\mathbf{L}_h=\mathbf{L}_0+\hat{\mathbf{L}}_h,
\]
where, with the notation of (\ref{hsplit}),
\begin{equation} 
\label{bhatL}
(\hat{\mathbf{L}}_hv)_i=\hat{h}^{jk}\partial_j\partial_k v_i
+\frac{1}{3}\hat{h}^{jk}\partial_i\partial_k v_j
+B^{jk}\,_i \partial_j\,v_k+
A^j\,_i\,v_j, 
\end{equation}
with
\[
B^{kj}\,_i= -2\,h^{jl}\,\Gamma_l\,^k\,_i -
\frac{4}{3}\,h^{lf}\,\Gamma_l\,^j\,_f\,h^k\,_i 
+\frac{1}{3}\,\partial_i\,h^{jk},
\]
and $A^j\,_i$ is a function of the metric coefficients and their
first and second derivatives. 
The fields $A^j\,_i$, $B^{kj}\,_i$ are smooth and
satisfy
\begin{equation} 
\label{BB1}
A^j\,_i = O(r), \quad B^{kj}\,_i=O(r^2), 
\end{equation}
and, because $x_k\,x^i\,\Gamma_i\,^k\,_j = 0$ at the point with normal
coordinates $x^k$, 
\begin{equation} 
\label{B2}
x_k\,x^i\,B^{kj}\,_i = - \frac{4}{3}\,r^2\,h^{lf}\,\Gamma_l\,^j\,_f.
\end{equation}
Similarly, we write the operator $\mathcal{L}_h$ in the form 
\[
\mathcal{L}_h=\mathcal{L}_0+\hat {\mathcal{L}}_h.
\]

\begin{lemma} 
\label{LhatY}
Suppose $p^i \in \mathcal{Q}_m$. Then $\hat{\mathbf{L}}_h (r^s
p^i)=r^{s-2}U^i$ with some $U^i \in \mathcal{Q}_\infty (B_a)$ which
satisfies $U^i =O(r^{m+3})$.
\end{lemma}
{\bf Proof:} Using (\ref{bhatL}), we  calculate 
$r^{-s+2}\,\hat{\mathbf{L}}_h(r^s p^i)$ and find
\begin{multline}
U_i= \hat{h}^{kj} \left(s \delta_{kj} p^i
+r^2 \partial_k \partial_j p_i \right)
+\frac{1}{3}\hat{h}^{jk}\left(  s x_i\partial_k p_j 
+r^2 \partial_k \partial_i p_j+\delta_{ki}p_j \right)\\
+B^{kj}\,_i \left(sx_k p_j+r^2 \partial_k p_i \right)
+r^2 A^j\,_i\,p_j.
\end{multline}
Thus $U^i$ is smooth. Using (\ref{BB1}) we obtain that 
$U_i=O(r^{m+3})$, using (\ref{B2}) we find $x^i U_i =r^2 f$
with some smooth function $f$. 
$\blacksquare$\\

We are in a position now to prove for $m=\infty$ the analogue of point
(ii) of theorem \ref{EL}.   

\begin{theorem} 
\label{estimateJ}
Assume that $h$ is smooth, $s\in \mathbb{Z}$, $s<0$, $|s|$ odd,
$F^i\in C^\infty (B_a)$, and $J^i \in \mathcal{Q}_\infty (B_a)$ with
$J^i=O(r^{s_0})$ for some $s_0 \ge |s|$.

Then, if $v^i \in W^{2,p}_{loc}(B_a)$ solves 
\[
(\mathbf{L}_h v)^i= r^{s-2} J^i+F^i,
\]
it can be written in the form 
\begin{equation} \label{vasymptotic}
v^i = r^s v^i_1 +v^i_2,
\end{equation}
with $v^i_1 \in \mathcal{Q}_\infty(B_a)$, $v^i_1=O(r^{s_0})$,  
$v^i_2 \in C^{\infty}(B_a)$.
\end{theorem}
{\bf Proof:}
The proof is similar to that of Theorem \ref{EL}. 
For given $m \in \mathbb{N}$ we can write by our assumptions
$J^i =T^i_{m}+J^i_R$, where $J^i_R = O(r^{m + 1})$ and  
$T^i_{m}$ denotes the Taylor polynomial of $J^i$ of order $m$.
Because $J^i \in  \mathcal{Q}_\infty (B_a)$, its Taylor polynomial can
be written in the form
\[
T^i_{m}=\sum_{k=s_0}^{m}t^i_k 
\quad \text{with} \quad t^i_k \in \mathcal {Q}_k.
\]
We define now a function $v^i_R$ (depending on $m$) by 
\[
v^i =r^s \sum_{k=s_0}^m v^i_k + v^i_R.
\]
The quantities $(v^i_k) \in \mathcal {Q}_k$ are determined by the
recurrence relation
\[
\mathbf{L}_0(r^s v^i_{s_0})=r^{s-2}t^i_{s_0}, \quad
\mathbf{L}_0(r^s v^i_k)=r^{s-2}(t^i_k-U^{(k)i}_{k}),
\]
where, for given $k$, the quantity $U^{(k)i}_{k} \in \mathcal {Q}_k$ is
obtained as follows. The function
\[
U^{(k)i} \equiv 
r^{-s+2}\,\hat {\mathbf{L}}(r^s \sum_{j=s_0}^{k-1} v^i_k),
\]
has by lemma \ref{LhatY} an expansion
\[
U^{(k)i}=\sum_{j=s_0+2}^m  U^{(k)i}_{j} + U^{(k)i}_R
\quad \text{with} \quad 
U^{(k)i}_{j} \in \mathcal {Q}_j, 
\]
from which we read off $U^{(k)i}_{k}$. By Lemma \ref{L0P} the recurrence
relation is well defined.

With these definitions, the remainder $v^i_R$ satisfies the equation
\[
\mathbf{L} v^i_R=r^{s-2} \left(  U^{(m+1)i}_R   +J^i_R  \right)+F^i.
\]
By lemma \ref{r-1} the right hand side of this equation is in
$C^{m+s-2,\alpha}(B_\epsilon)$. By elliptic regularity we have
$v^i_R \in C^{m+s,\alpha}(B_a)$. Since $m$ was arbitrary, the conclusion
follows now by an argument similar to the one used in the proof of lemma 
\ref{Einfty}.
$\blacksquare$\\ 

Theorem \ref{estimateJ} will allow us to prove that the solutions of
the momentum constraint obtained in section \ref{existencemom} have an 
expansion of the form (\ref{PsiS}), if we impose near $i$ certain
conditions on the data which can be prescribed freely. 

In definition (\ref{eq:Qn}) of the field $\Phi^{ab}_{sing}$, which 
enters theorem (\ref{sing}), we assume that $\Psi^{ab}_{flat}$ is of
the form (\ref{gensol}) with $\lambda \equiv 0$, i.e. it is given by the
tensor fields (\ref{eq:PsiPp})--(\ref{eq:PsiQp}).
In order to write $\Psi^{ab}_{flat}$ in a convenient form, we introduce 
vector fields which are given in normal coordinates by
\begin{xalignat}{2}
v_P^i & = -\frac{1}{4} P^k\partial_k \partial^i r^{-1}  
= r^{-5}p_P^i, & \quad  
p_P^i & =\frac{1}{4}(r^2\,P^i
-3\,x^i\,P^k\,x_k)\in \mathcal{Q}_2, \label{eq:PvP}\\
v_J^i & =\epsilon^{ijk}J_j\partial_k r^{-1} = r^{-3}p_J^i, & \quad
p_J^i & = - \epsilon^{ijk}\,J_j\,x_k\in \mathcal{Q}_1, \label{eq:PvvJ}\\
v_A^i & = \frac{1}{2}\,A\,\partial^i r^{-1} = r^{-3}p_A^i, & \quad  
p_A^i & = -\frac{1}{2}\,A\,x^i\in \mathcal{Q}_1, \label{eq:PvA}\\  
v_Q^i & = -2\,Q^i r^{-1}
+\frac{1}{4} Q^k\partial_k\partial^i r = r^{-3}p_Q^i, & \quad  
p_Q^i & = - \frac{7}{4}\,r^2 Q^i 
- \frac{1}{4}\,x^i\,Q^k\,x_k \in \mathcal{Q}_2,\label{eq:PvQ}
\end{xalignat}
where $P^i$, $J^i$, $A$, $Q^i$ are chosen such that the vector
fields satisfy
\begin{equation}
\label{eq:vPsi}
(\mathcal{L}_0 v_P)^{ab}=\Psi_P^{ab}, \quad 
(\mathcal{L}_0 v_J)^{ab}=\Psi_J^{ab}, \quad
(\mathcal{L}_0v_Q)^{ab}=\Psi_Q^{ab}, \quad 
(\mathcal{L}_0 v_A)^{ab}=\Psi_A^{ab},    
\end{equation}
with $\Psi_P^{ab}$, $\Psi_J^{ab}$, $\Psi_A^{ab}$, $\Psi_Q^{ab}$ 
as given by (\ref{eq:PsiPp})--(\ref{eq:PsiQp}). We have 
on $B_a \setminus \{i\}$
\begin{equation}
\label{eq:vL0}
(\mathbf{L}_0 v_P)^a = 0, \quad
(\mathbf{L}_0 v_J)^a = 0, \quad
(\mathbf{L}_0 v_A)^a = 0, \quad
(\mathbf{L}_0 v_Q)^a = 0,  
\end{equation}
and can thus write on $B_{a/2} \setminus \{i\}$
\begin{equation} 
\label{eq:Qv}
\Phi^{ij}_{sing} = 
(\mathcal{L}_{0} (v_P + v_J + v_A + v_Q))^{ij}
- \frac{1}{3}\,h^{ij}\,h_{kl}\,
(\mathcal{L}_{0} (v_P + v_J + v_A + v_Q))^{kl}.
\end{equation}

Of the field $\Phi^{ab}_{reg} \in W^{1,p}(S)$, $p > 1$, entering theorem
\ref{sing} we assume that it can be written near $i$ in the form
\begin{equation}
\label{eq:conregQ}
\Phi^{ab}_{reg} = r^{s}\,\Phi^{ab}_{1 reg} + \Phi^{ab}_{2 reg}, 
\end{equation}
where $s \le - 1$ is some integer which will be fixed later on,
$\Phi^{ab}_{1 reg}$, $\Phi^{ab}_{2 reg}$ are smooth in $B_a$
and such that $\Phi^{ij}_{1 reg} = O(r^{-s-1})$,
and $x_i\,x_j\,\Phi^{ij}_{1 reg} = r^2\,\Phi$ with some 
$\Phi \in C^{\infty}(B_a)$. Then 
\begin{equation}
\label{eq:JQ}
J_{reg}^i \equiv D_j\,\Phi^{ij}_{reg} = r^{s-2} \hat J^i 
+ D_j\,\Phi^{ij}_{2 reg},
\end{equation}
with $\hat J^i=r^2\,D_j\,\Phi_{1 reg}^{ij} 
+ s\,x_j\,\Phi_{1 reg}^{ij} \in \mathcal{Q}_\infty$ 
and $\hat J^i=O(r^{-s})$.\\ 

Using theorem \ref{estimateJ}, we obtain for the solutions of theorem
\ref{sing}  (where we can set by our present assumptions 
$\omega_0 \equiv 1, h \equiv h'$) the following result.

\begin{corollary}
\label{linealmestimate}
With the tensor fields $\Phi^{ab}_{sing}$, $\Phi^{ab}_{reg}$ given by 
(\ref{eq:Qv}), (\ref{eq:conregQ}) respectively, let the vector field
$v^a$ be such that
\begin{equation}
\label{phisqcontr}
\Psi^{ab} = \Phi^{ab}_{sing} + \Phi^{ab}_{reg} + (\mathcal{L}v)^{ab},
\end{equation}
satisfies $D_a\,\Psi^{ab}=0$ in $B_a \setminus \{i\}$.\\

\noindent
(i) If $P^a = 0$ in (\ref{eq:Qv}) and $s=-3$ in (\ref{eq:conregQ}),
the vector field $v^a$ can be written in the form
\[
v^i=r^{-3}\,v_1^i + v_2^i, \quad \text{with} \quad 
v^i_1 \in \mathcal{Q}_\infty(B_a), \quad v^i_1=O(r^{3}),
\quad v^i_2 \in C^{\infty}(B_a).
\] 

\noindent
(ii) If $J^a = 0$, $A = 0$, $Q^a = 0$ in (\ref{eq:Qv}) and $s=-5$ in
(\ref{eq:conregQ}), the vector field $v^a$ can be written in the form
\[
v^i=r^{-5}\,v_1^i + v_2^i, \quad \text{with} \quad 
v^i_1 \in \mathcal{Q}_\infty(B_a), \quad v^i_1=O(r^{5}),
\quad v^i_2 \in C^{\infty}(B_a).
\] 
\end{corollary}
{\bf Proof:}
In both cases the vector field $v^a$ satisfies
$\mathbf{L}_h v^a = -J^a_{sing} - J^a_{reg}$ with  
$J^a_{reg}$ given by (\ref{eq:JQ}) and 
$J^a_{sing} = D_b\,\Phi^{ab}_{sing}$. 
By equation (\ref{eq:vL0}) we have in case (i) 
$J^a_{sing} = (\mathbf{L}_h (v_J+v_A+v_Q))^a = (\mathbf{\hat L}_h
(v_J+v_A+v_Q))^a$, and in case (ii)  
$J^a_{sing} = (\mathbf{L}_h\,v_P))^a = (\mathbf{\hat L}_h\,v_P))^a$ on
$B_{a/2} \setminus \{i\}$. The
results now follow from equations (\ref{eq:PvP})--(\ref{eq:PvQ}),
lemma \ref{LhatY}, and theorem \ref{estimateJ}. 
$\blacksquare$\\

We are in a position now to describe the behaviour of the scalar field
$\Psi_{ab}\,\Psi^{ab}$ near $i$.

\begin{lemma} 
\label{PsiPsi}
The tensor field (\ref{phisqcontr}) satisfies\\

\noindent
in case (i) $\,\,r^8\,\Psi_{ab}\,\Psi^{ab} \in E^\infty(B_a)$,\\

\noindent
in case (ii)
$\,\,r^8\,\Psi_{ab}\,\Psi^{ab} =\psi +r\,\psi^R$
where $\psi^R\in C^{\alpha}(B_a)$ and
$\psi = \frac{15}{16}\,P_i\,P^i +r^{-2}\,h_2$ with harmonic polynomial
$h_2 = \frac{3}{8}\,r^2\,(3\,(P^i\,n_i)^2 - P_i\,P^i)$.
\end{lemma}
{\bf Proof:}
For $w^i \in \mathcal{Q}_\infty(B_a)$ with $x_i\,w^i = r^2\,\hat{w}$, 
$\hat{w} \in C^{\infty}(B_a)$ we have
\begin{equation}
\label{eq:calLv1}
( \mathcal{L}_h(r^s\,w))^{ij}= r^s\,((\mathcal{L}_h w)^{ij} 
- \frac{2}{3}\,s\,h^{ij}\,\hat{w}) + r^{s - 2}\,2\,s\,x^{(i}\,w^{j)}. 
\end{equation}
We set $s = - 3$ and $w^i = p^i_J + p^i_A + p^i_Q$ in case (i) and 
$s = - 5$ and $w^i = p^i_P$ in case (ii) and we write 
$x_i\,v^i_1 = r^2\,\hat{v}_1$ with $\hat{v}_1 \in C^{\infty}(B_a)$.  
Observing the equation above we get on $B_{a/2} \setminus \{i\}$ a
representation 
\[
\Psi^{ij} = r^s\,H^{ij} + r^{s - 2}\,K^{ij} + L^{ij}
\]
with fields
\[
H^{ij} = (\mathcal{L}_0\,w)^{ij} - \frac{2}{3}\,h^{ij}\,
h_{kl}\,(\mathcal{L}_0\,w)^{kl} 
- \frac{2}{3}\,s\,\hat{w}\,\left(\delta^{ij} 
+ 2\,h^{ij}\,(1 - \frac{1}{3}\,h_{kl}\,\delta^{kl})\right)
\]
\[
+ (\mathcal{L}_h\,v_1)^{kl} - \frac{2}{3}\,s\,h^{ij}\,\hat{v}_1
+\Phi^{ij}_{1 reg},
\]
\[
K^{ij} = 2\,s\,(x^{(i}\,w^{j)} + x^{(i}\,v_1^{j)}), \quad
L^{ij} = \Phi^{ij}_{2 reg} + (\mathcal{L}_h\,v_2)^{ij},
\]
which are in $C^{\infty}(B_{a/2})$. Since a direct calculation gives
$K_{ij}\,K^{ij} = r^2\,K$ with $K \in C^{\infty}(B_{a/2})$, we get
\[
\Psi_{ij}\,\Psi^{ij} = 
r^{2 s - 2}\,(2\,H_{ij}\,K^{ij} - K) 
+ r^{2 s}\,H_{ij}\,H^{ij}
\]
\[
+ r^{s - 2}\,2\,K_{ij}\,L^{ij} 
+ r^s\,2\,H_{ij}L^{ij}
+ L_{ij}\,L^{ij},
\]
from which we can immediately read off the desired result in case (i).
In case (ii) it is obtained from our assumptions by a detailed calculation
of $r^{2 s - 2}\,(2\,H_{ij}\,K^{ij} - K)  + r^{2 s}\,H_{ij}\,H^{ij}$.
$\blacksquare$\\

Combining the results above and observing the conformal invariance of the
equations involved, we obtain the following detailed version of theorem
\ref{T2}. We use here the notation of theorem \ref{sing}.

\begin{theorem}
Assume that the metric $h$ is smooth and $\Psi^{ab}$ is the 
solution of the momentum constraint determined in theorem \ref{sing}. 
If\\ 

\noindent
(i) $\Phi^{ab}_{sing}
= \Psi_J^{ab} + \Psi_A^{ab} + \Psi_Q^{ab} 
- \frac{2}{3}\,h^{ab}\,h_{cd}\,(\Psi_J^{cd} + \Psi_A^{cd} + \Psi_Q^{cd})$
in $B_{a/2}$,\\

\noindent
(ii) $\Phi^{ab}_{reg} = r^{-3}\,\Phi^{ab}_{1 reg} + \Phi^{ab}_{2 reg}$
with $\Phi^{ab}_{1 reg}, \Phi^{ab}_{2 reg} \in C^{\infty}(B_a)$
such that $\Phi^{ab}_{1 reg} = O(r^2)$,
and $x_a\,x_a\,\Phi^{ab}_{1 reg} = r^2\,\Phi$ with some 
$\Phi \in C^{\infty}(B_a)$,\\

\noindent
then $\Psi^{ab}$ satisfies condition (\ref{condd}). 
\end{theorem}

\begin{appendix}

\section{On H\"older functions} 
\label{holder}

In this section we want to proof an estimate concerning H\"older
continuous functions.

Let $B$ be an open ball in $\mathbb{R}^{n}$, $n\ge 1$, centered at the
origin. Suppose $f \in C^k(U)$ for some $k \ge 0$ and $m$ is a
non-negative integer with $m \le k$. Then we can write
\[
f = 
\sum_{|\beta| < m} \frac{1}{\beta !}\,\partial^{\beta}f(0)\,x^{\beta}
+ m\,\int_0^1 (1 - t)^{m - 1}\sum_{|\beta| = m} 
\frac{1}{\beta !}\,\partial^{\beta}f(t\,x)\,x^{\beta}\,d\,t
\]
\[
=
\sum_{|\beta| \le m} \frac{1}{\beta !}\,\partial^{\beta}f(0)\,x^{\beta}
+ m\,\int_0^1 (1 - t)^{m - 1}\sum_{|\beta| = m} 
\frac{1}{\beta !}\,
(\partial^{\beta}f(t\,x) - \partial^{\beta}f(0))\,x^{\beta}\,d\,t,
\]
where the first line is a standard form of Taylor's formula and
the second line a slight modification thereof. We denote by $T_m(f)$ the
Taylor polynomial of order $m$ and by $R_m(f)$ the modified remainder,
i.e. the first and the second term of the second line respectively.  

\begin{lemma} 
\label{holderrest}
Suppose $f \in C^{m,\alpha}(U)$. Then  
$f - T_m(f) \in C^{m,\alpha}(U)$ and we have for 
$\beta \in \mathbb{N}_0^n$, $|\beta| \le m$, 
\begin{equation}
\label{hineq}
|\partial^\beta (f - T_m(f))(x)| \leq 
|x|^{m + \alpha - |\beta|}\,\sum_{|\gamma| = m - |\beta|}
\frac{1}{\gamma !}\,c_{\gamma + \beta}
\quad\mbox{on}\quad U,
\end{equation}
where the constants $c_{\delta}$ denote the  H\"older coefficients 
satisfying
$|\partial^{\delta}f(x) - \partial^{\delta}f(0)| \le
c_{\delta}\,|x|^{\alpha}$ in $U$ for $\delta \in \mathbb{N}_0^n$, 
$|\delta| = m$.    
\end{lemma} 

{\bf Proof:}
Applying the modified Taylor formula to $f$ and then to its derivatives,
we get
\[
\partial^{\beta}(f - T_m(f)) =
T_{m - |\beta|}(\partial^{\beta} f) 
+ R_{m - |\beta|}(\partial^{\beta} f) 
- \partial^{\beta} T_m(f).
\]
We show that 
\begin{equation}
\label{Tder=derT}
T_{m - |\beta|}(\partial^{\beta} f) - \partial^{\beta} T_m(f) = 0.
\end{equation}
To prove this equation we use induction on $n$.  For $n = 1$ the result
follows by a direct calculation. To perform the induction step we assume
$n \ge 2$ and show that the statement for $n - 1$ implies that for $n$.

We write 
$x = (x', x^n)$ for $x \in \mathbb{R}^n$ and 
$\beta = (\beta', \beta_n)$ for $\beta \in \mathbb{N}_0^n$ etc. Then
we find the equalities
\[
\partial^{\beta}T_m(f) = 
\partial^{\beta'}\partial^{\beta_n}
(\sum_{\gamma_n = 0}^m T_{m - \gamma_n}(\partial^{\gamma_n}f)
\,\frac{1}{\gamma_n !} (x^n)^{\gamma_n})
\]
\[
= \sum_{\gamma_n = \beta_n}^{m - |\beta| + \beta_n}
T_{m - |\beta'| - \gamma_n}(\partial^{\beta'}\partial^{\gamma_n}f)
\frac{1}{(\gamma_n - \beta_n)!}(x^n)^{\gamma_n - \beta_n}
\]

\[
= \sum_{\gamma_n = 0}^{m - |\beta|}
T_{m - |\beta| - \gamma_n}(
\partial^{\beta'}\partial^{\gamma_n + \beta_n}f)
\frac{1}{\gamma_n!}(x^n)^{\gamma_n} =
T_{m - |\beta|}(\partial^{\beta} f).
\]
Here the first line is a simple rewriting where we denote by 
$T_{m - \gamma_n}(\partial^{\gamma_n}f)$ the Taylor polynomial of order 
$m - \gamma_n$ of the function $\partial^{\gamma_n}f(x', 0)$ of 
$n - 1$ variables. In the second line the
derivatives are taken and the induction hypothesis is used. The
third line is obtained by redefining the index $\gamma_n$ and  
using a similar rewriting as in the first line.

With (\ref{Tder=derT}) the estimate (\ref{hineq}) follows  immediately by 
estimating the integral defining  $R_{m - |\beta|}(\partial^{\beta}f)$.
$\blacksquare$

\section{An additional result} 
\label{additional}
In this section we prove a certain extension of theorem \ref{EL}.

\begin{theorem} 
\label{EmSchauder}
Let $u$ be a distribution satisfying $Lu=f$, where 
$f\in E^{m,\alpha}(B_a)$, and the coefficient of the elliptic operator $L$
are in $C^{m,\alpha}(B_a)$. Then 
\begin{equation}
\label{ur3exp}
u= r^3\sum_{k=0}^m u_k +u_R \in E^{m+2,\alpha}(B_a),
\end{equation}
with $u_k \in \mathcal{P}_k$ and $u_R \in C^{m+2,\alpha}(B_a)$.
\end{theorem}
{\bf Proof:}
We follow the proof of Theorem \ref{EL} using Schauder instead of $L^p$
estimates. Since $f = f_1 + r f_2 \in E^{m,\alpha}(B_a)$ we have 
\[
f =r\,T_m + f_R
\quad\mbox{with}\quad
T_m=\sum_{k=0}^m t_k,
\]
where $T_m$ is the Taylor polynomial of order $m$ of $f_2$ and 
$t_k \in \mathcal{P}_k$. 

Consider the recurrence relation 
\[
\Delta(r^3u_0)=rt_0,\,\,\,\,\,\,
\Delta(r^3u_k)=r(t_k-U^{(k)}_k),\,\,\,\,\,\,
1 \le k \le m,
\]
which is obtained by defining $U_k$, $k = 1, \ldots, m$, by 
\[
\hat L(r^3 \sum_{j=0}^{k-1}u_j)=r U^{(k)},
\]
and defining $U^{(k)}_j$ and $U^{(m + 1)}_j$ as in the proof of theorem 
\ref{EL}. The equation for $u$ and (\ref{ur3exp}) then imply for $u_R$ 
equation (\ref{fuR}) with $s = 3$. Since by our assumptions and lemma \ref{r-1}
the right-hand side of this  equation is in $C^{m,\alpha}(B_a)$,
the interior Schauder estimates of theorem \ref{SchauderEstimate} imply that 
$u_R \in C^{m+2}(B_a)$. 
$\blacksquare$

\end{appendix}


\end{document}